\documentclass[preprint,5p,times]{elsarticle}

\usepackage[utf8]{inputenc} 
\usepackage[T2A]{fontenc}
\usepackage[english]{babel}   
\usepackage{lineno}
\usepackage{graphicx}
\usepackage{xcolor}
\usepackage[pdftex, backref, colorlinks]{hyperref}
\usepackage{multirow}
\usepackage{enumitem}

\journal{Universe}
\begin{document}


\begin{frontmatter}
\title{EAS observation conditions in the SPHERE-2 balloon experiment}

\author[address1]{E.A.~Bonvech\corref{correspondingauthor1}}
\cortext[correspondingauthor1]{Corresponding author}
\ead{bonvech@yandex.ru}
\author[address1]{D.V.~Chernov}
\author[address2,address3]{M.~Finger}
\author[address2,address3]{M.~Finger Jr.}
\author[address4]{V.I.~Galkin}
\author[address4,address1]{D.A.~Podgrudkov\corref{correspondingauthor1}}
\ead{d.a.podgrudkov@physics.msu.ru}
\author[address1]{T.M.~Roganova}
\author[address4,address1]{I.A.~Vaiman}
\address[address1]{M.V. Lomonosov Moscow State University, Skobeltsyn Institute of Nuclear Physics (SINP MSU), Moscow, Russia}
\address[address2]{Charles University, Faculty of Mathematics and Physics, Prague, Czech Republic}
\address[address3]{Joint Institute for Nuclear Research, Dubna, Russian Federation}
\address[address4]{M.V. Lomonosov Moscow State University, Faculty of Physics, Moscow, Russia}

\begin{abstract}
The SPHERE project studies primary cosmic rays by detection of the Cherenkov light of extensive air showers reflected from the snowy surface of the earth. Measurements with the aerial-based detector SPHERE-2 were performed in 2011--2013. The detector was lifted by the balloon at altitudes up to 900~m above snowed surface of Lake Baikal, Russia. The results of the experiment are summarized now in a series of papers that opens with this article.

An overview of the SPHERE-2 detector telemetry monitoring systems is presented along with the analysis of the measurements conditions including atmosphere profile. The analysis of the detector state and environment atmosphere conditions monitoring provided various cross-checks of detector calibration, positioning and performance.
\end{abstract}

\begin{keyword}
primary cosmic rays\sep extensive air showers\sep Vavilov-Cherenkov radiation\sep balloon \sep reflected Cherenkov light
\MSC[2010] 00-01\sep  99-00
\end{keyword}
\end{frontmatter}

\section{Introduction}
The SPHERE-2 experiment was designed for primary cosmic ray (PCR) studies in the 10--1000~PeV energy range. PCR particles induce secondary particle cascades named extensive air showers (EAS) and secondary radiations such as Cherenkov light, fluorescent light, radio emission etc. in the atmosphere that can be registered by different methods by ground-based detectors. 
The SPHERE project is the first successful implementation of a new EAS detection method --- detection of reflected Cherenkov light (CL) using an aerial-based detector --- a method first proposed by A.~Chudakov~\cite{chu74:VKKL74} and first implemented by R.~Antonov~\cite{ant75, ant86, ant97, Ant15a}.

Common EAS arrays such as the Telescope Array~\cite{abu12}, Yakutsk EAS array~\cite{Yakutsk19} or the TAIGA~\cite{TAIGA20} detector are ground-based structures spread over an area of up to several hundreds~\cite{abu12} square kilometers. Significant efforts are required to install the sensitive elements of such vast arrays and to maintain their network connected, powered and time-synchronized. Even more effort is required for regular calibration of detector stations and atmosphere parameter monitoring over the entire detector area. 

On the other hand, conventional imaging air Cherenkov telescopes (IACT), like HESS~\cite{HESS03a, HESS03b} or MAGIC~\cite{MAGIC16-1, MAGIC16-2} are relatively compact systems that have good calibration means, high integrity power supply, atmosphere transparency control, and so on. But since the CL from EAS has a very narrow directional pattern it cannot be observed by detectors located at a distance of more than 0.5--1.5~km from the shower axis. Therefore, IACTs have a relatively low upper energy threshold.

The method of detection of the reflected CL allows, on one hand, to register EAS on a relatively large area and later reconstruct the lateral distribution function of Cherenkov photons, and, on the other hand, to utilize a small size compact detector with all the advantages of such a setup. The mentioned advantages include, but are not limited to, the opportunity to implement: complex topological trigger conditions prior to writing data to storage thus increasing the maximum operational count rate; direct on-line calibration system; high mobility with lower operational costs and etc.

A compact detector that observes large surface areas, however, can combine some strong sides of an IACT with a higher upper energy threshold. The main drawback of this design, however, is that the detector is significantly limited in its weight and therefore size. This means that the detector's sensitive element should preferably operate in photon counting mode. This requires a good understanding of the detector state and measuring conditions.

The general overview of the SPHERE experiment can be found in~\cite{Ant15a}, and a detailed description of the SPHERE-2 electronics is given in~\cite{Ant20}.

Here we present an overview of the data obtained by various auxiliary sensors on board of the SPHERE-2 detector. This data, collected during and in some cases before and after EAS measurements, was used in the subsequent analysis for detector state and environment conditions monitoring, allowing to perform various cross-checks of detector calibration, positioning and performance.

\section{The SPHERE-2 detector telemetry systems}
\label{sect:detector}
The \mbox{SPHERE-2}~\cite{Ant20} is a compact detector designed to register EAS CL reflected from the earth snow. The detector was lifted using a tethered balloon to altitudes of up to 0.9~km above the snow covered surface. A special tethered balloon BAPA~\cite{Ant20} was developed and manufactured by the Russian Augur balloon systems agency~\cite{Augur}.

\begin{figure}[bt]
\centering
    \includegraphics[width=0.42\textwidth]{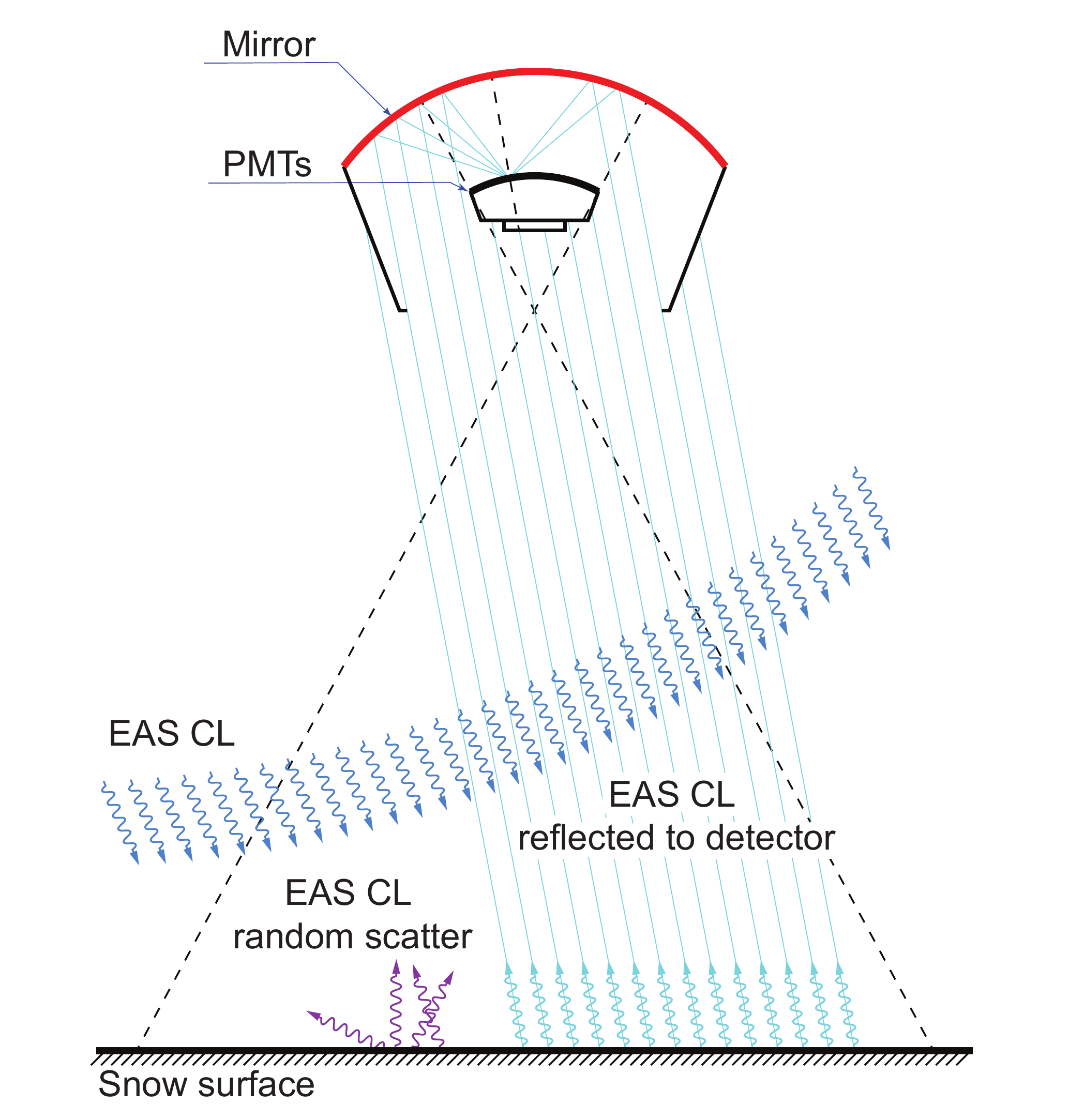}
    \caption{The detector optical scheme.}
\label{fig:optics}
\end{figure}

The \mbox{SPHERE-2} detector optics were comprised of a 1.5~m diameter spherical mirror with a 109 photomultiplier tubes (PMT) mosaic. The optical scheme of the detector is shown in Fig.~\ref{fig:optics}.

The PMT mosaic was located near the mirror focal surface and recorded CL reflected from the snow surface below the detector. The PMTs were arranged in a hexagonal structure on the spherical surface, see details in~\cite{Ant20}, or below on Fig.~\ref{fig:2012-3_shore_image}. FEU-84/3 PMTs were used in the mosaic. The exact numbers of PMTs in each season are given in Tab.~\ref{tab:statistics}. The fields of view of the PMTs are fixed in angular terms. The exact shape of the area that a single PMT observes depends on the PMT's position in the mosaic and the detector orientation.

\begin{table}[tb]
\centering
\caption{The annual statistics of the SPHERE experiment on Baikal Lake.
}
\label{tab:statistics}
\vspace{1pc}
\begin{tabular}{|c||c|c|c|r|r|}
    \hline
    year  & flights & PMT    & time, & triggers \\ 
          &         & number & hours & detected \\ 
    \hline \hline
    \multicolumn{5}{|c|}{test runs} \\
    \hline
    2008 & 1 &  20 &  1 &  6180 \\ 
    2009 & 3 &  64 & 13 & 10312 \\ 
    2010 & 6 &  96 & 30 &  1343 \\
    \hline
    \multicolumn{5}{|c|}{experiment runs} \\
    \hline
    2011 & 4 &  96 & 30 & 20571 \\
    2012 & 5 & 108+1 & 31 &  7716 \\
    2013 & 5 & 108+1 & 33 &  3813 \\
    \hline
\end{tabular}
\end{table}

\subsection{Control block telemetry}

The control block of the SPHERE-2 detector contained a large set of various sensors, as well as the main recording electronics and a power supply system. Sensors located both inside and outside of the control block were used to monitor the detector state and performance and to collect supplementary data for EAS measurements. The list of sensors used, their measuring range, precision and data collection frequency is given in the Tab.~\ref{tab:telemetry_sensors}. 

\begin{table*}[bth]
\centering
\caption{SPHERE-2 telemetry data readout intervals and sensors precision.}
\label{tab:telemetry_sensors}

\begin{tabular}{|c|l|l|c|r@{\hspace{1mm}}c@{\hspace{1mm}}l|c|}
\hline
\multicolumn{1}{|c|}{interval} & \multicolumn{1}{c|}{parameter} & \multicolumn{1}{c|}{data}  & \multicolumn{1}{|c|}{accuracy} & \multicolumn{3}{c|}{range}  & \multicolumn{1}{c|}{units} \\
\hline
\multirow{6}{*}{1 sec} & \multirow{3}{*}{Detector position} &GPS altitude & (1)* &  -1500&--&18000  & m a.s.l.\\
                                                      \cline{3-8}
                         &                              & GPS position & 4 (2)** & &---&& m\\
                                                      \cline{3-8}
                       &                              & GPS time (PPS)& 1 & &---&& $\mu$s \\
                       \cline{2-8}
                       & \multirow{2}{*}{Detector orientation} & inclination angles (resolution) X,Y& 0.3 (0.02)&$-$25&--&25&deg\\
                                                      \cline{3-8}
                       &                              & compass azimuth (resolution) Z &2.5 (0.5)&0&--&360&deg\\
                       \cline{2-8}
                       &Control block                 & inner temperature& 1.5 & $-40$&--&70 &deg,$^\circ$C\\
\hline
\multirow{7}{*}{1 min} & \multirow{2}{*}{PMT status} & anode current & 0.03 & 0&--&125 & $\mu$A\\
                                                      \cline{3-8}
                       &                              & mosaic temperature & 1.5 & $-40$&--&70 & deg,$^\circ$C\\
                       \cline{2-8}
                       & Power source                 & high voltage (HV1) & 0.1 & 0&--&250 & V\\
                       \cline{2-8}
                       & \multirow{2}{*}{Barometer}   & pressure & 5 & 750&--&1100 & hPa\\
                                                      \cline{3-8}
                       &                              & temperature& 2 & -20&--&60 &deg,$^\circ$C\\
                       \cline{2-8}
                       & Balloon barometer            & pressure   & 3 & 0&--&1000 & Pa\\
                       \cline{2-8}
                       & Battery (19V)                & voltage & 0.01 & 0&--&40 & V\\
                       \cline{2-8}
                       & Constant voltage (5V)        & voltage & 0.01 & 0&--&40 & V\\
\hline
\multirow{5}{*}{10 min} & \multirow{3}{*}{PMT status} & first dinode voltage & 0.06 & 0&--&250 & V\\
                                                      \cline{3-8}
                       &                              & PMT temperature & 0.1 & -30 &--&50 & deg,$^\circ$C\\
                                                      \cline{3-8}
                       &                              & supply voltage & 0.006 & 0&--&25 & V\\
                       \cline{2-8}
                       & Trigger                      & counting rate &1&0&--&4& Hz\\
                       \cline{2-8}
                       & FADC boards                  & voltage (1.2, 2.5, 2.8) & 0.001 & 0&--&4 & V\\
\hline
\end{tabular}

\vspace{1mm}

\footnotesize \raggedright 
\hspace{5 mm}* 
1 m accuracy according to our own analysis (see in section~\ref{sect:orientation} )

\hspace{5 mm}**  3–5 meters, 95\% typical accuracy according to the manufacturer provided data, 2 m accuracy according to our own analysis (see in section~\ref{sect:orientation} )
\normalsize
\end{table*}


The sensors inside the control block measured:
\begin{itemize}[nosep]
\item temperature of the FADC boards by two sensors per board for cooling system operation;
\item voltage on the secondary power supply units for PMT mosaic power stabilization;
\item voltage and current on the main power supply unit;
\item overpressure and temperature inside the balloon.
\end{itemize}
The sensors outside the control block measured:
\begin{itemize}[nosep]
\item detector position (GPS-module);
\item local air pressure and temperature;
\item horizontal orientation (compass).
\end{itemize}
Additional sensors were installed on the PMTs in the mosaic and controlled
\begin{itemize}[nosep]
\item power supply voltage;
\item anode current;
\item temperature.
\end{itemize}

Some of the data provided by these sensors was used to monitor the flight conditions and detector status in real time. Also, all of the data was stored and later used in EAS parameters reconstruction and detector performance evaluation.

\subsection{Orientation control system}
\label{sect:orientation}

At the rest state the detector axis was oriented vertically and the detector observed the snow covered surface just under itself. During the measurements the detector was hung under the balloon and swung and rotated freely. Its inclination is an important factor for shower parameter reconstruction since it determines the overall geometry of the experiment. To control the position and rotation of the detector a dual-axis inclinometer and a digital compass were installed on the PMT mosaic control board. 

The compass measured the angle of the detector's orientation relative to the Earth's magnetic field with a 2.5~degree precision.

The inclinometer measured two angles between its internal orthogonal axes and the plane orthogonal to the gravity vector. The orientation of the inclinometer's axes respectively to detector mosaic axes was determined in the laboratory. After careful preparations the mosaic was set horizontally with precision better than $0.5^\circ$ and the zero level of the inclinometer was measured. All detector inclination angles were calculated considering the recovered zero level.

The type of inclinometer used allows fast angles measurements with low power consumption, but, unlike the gyroscopic ones, it is affected by acceleration. As we will see in sec~\ref{sect:telemetrydata} during the experimental measurements there was recorded only a single instance of strong wind gusts at the end of the 2013-3 flight, and the balloon position was relatively stable at all other times. So we assumed that the registered inclination angles were unaffected by swinging.

The Garmin 16xHVS~\cite{GPS-module-specs} GPS sensor, located on the control unit was used to measure geographical coordinates and elevation above sea level. 

The GPS manufacturer declared a 3–5 meters, 95\% typical accuracy in position measurements.
Our observations show than the accuracy of position detection by the GPS sensor differed for moving and stationary situations. 

The accuracy of the GPS sensor in stationary state was calculated from GPS data when the detector was stationary on the ice. Daily distributions of the detector position and altitude were plotted for 20 days in 2012 and 2013. The median sigma of these distributions was taken as the accuracy of the position and altitude measurements: 2~m and 1~m respectively.

However, the detector showed a significant lag in altitude measurements during the initial balloon climb and subsequent flight altitude changes. More details on this lag, its correction and influence on detector altitude determination are discussed in Sec.~\ref{sect:gps_correction}.


\subsection{Telemetry monitoring\label{sect:telemetry}}

During every experimental flight a lot of information about the detector status and environmental parameters was recorded. Various telemetry data was received and checked every second, every minute and every 10 minutes. 

Every second the detector position and orientation was recorded. The GPS data (altitude, latitude and longitude), universal global time UTC, the angles of detector inclination, digital compass data and the control block inner temperature was monitored. An additional GPS and barometer were located in the control booth at ice level.

Every minute the PMT and power status were monitored: anode currents of each PMT, the PMT mosaic temperature, the temperature of the PMTs high voltage power source, power supply voltage and DC voltage sources were recorded. Barometer sensors were also polled every minute.

Every ten minutes the detector logged the PMTs supply voltages, voltages at PMTs first and tenth dynodes, individual PMT temperatures, the measuring channel counting rates and voltage on the FADC boards.

The PMTs power supply data was later used to control the operational stability of the optical modules. The registered discriminator triggering rate was used for discriminator thresholds selection in experimental flights. Deviations in power consumption by FADC boards was used to monitor possible malfunctions in the detector electronics. Data from temperature sensors was used for in-flight detector state and cooling system control.

\section{Experimental conditions}
\label{sect:data}
 
The SPHERE experiment was carried out on the Baikal Lake during winter seasons of 2008--2013. The Baikal lake ice thickness reaches 40--60~cm in February and can hold heavy vehicles. The balloon launch pad was deployed on the lake surface around 700~m from the shore (near 51$^\circ$\,47'\,48''~N, 104$^\circ$\,23'\,19''~E).

The \mbox{SPHERE-2} detector was lifted by a tethered balloon BAPA to altitudes of up to 900~m above the snow surface. Annual statistics of the experiment are given in Tab.~\ref{tab:statistics}. The first two years were dedicated to test runs. The detector configuration was improved from year to year. Thus the number of PMTs was increased from 96 to 109, and the signal sampling rate was changed in 2012 from~40~MHz to 80~MHz.


\subsection{Weather and snow}

Measurements were carried out during clear moonless nights with low wind conditions. The measurements were started 1.5 hours after sunset or immediately after moon set and finished before moon rise or 1.5 hours before sunrise. If the wind conditions became unsuitable for the flight of the balloon, the detector was landed at once. However, this happened only once during the third flight in 2013. The typical night air temperature on the lake surface was near $-15^\circ$C. According to the \href{https://rp5.ru/Weather_in_the_world}{Reliable Prognosis} data archive~\cite{rp5} from the nearest weather stations 30818 and 30710 the horizontal visibility was `at least 10 km' (the highest possible grade in the system), and the altitude of the base of the lowest clouds was `2500~m or more or no clouds present'.

\begin{figure}[tb]
\begin{center}
    \includegraphics[trim=1cm 5cm 0cm 5cm,clip,width=0.45\textwidth]{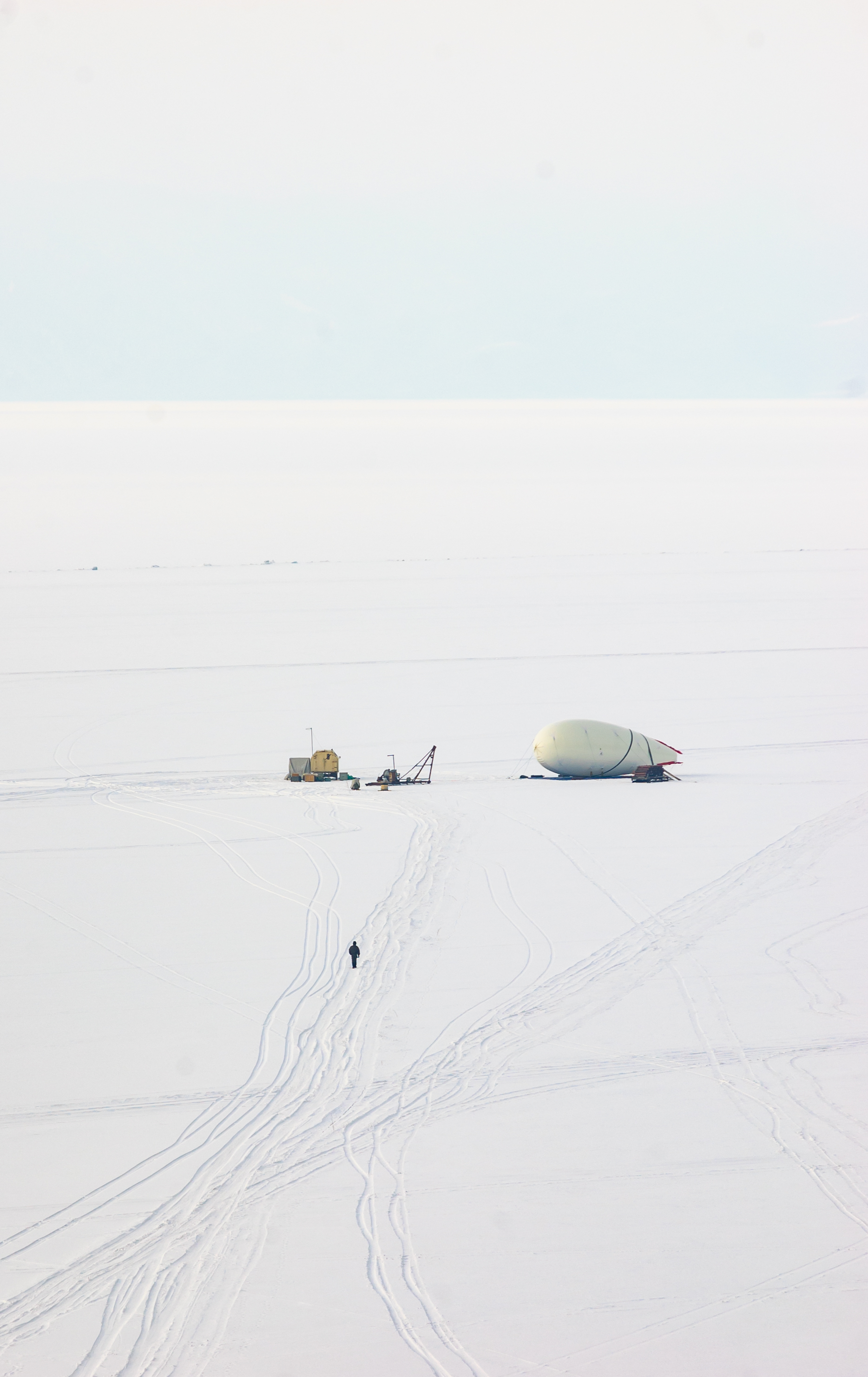}\hspace{2pc}%
    \caption{An overview of the launch site on the Baikal Lake from a hill on the shore. The snow coverage around the SPHERE start point was thick and usually renewed between flights by occasional snowfalls.}
\label{fig:baikal_snow}
\end{center}
\end{figure}

\begin{figure}[tb]
\begin{center}
    \includegraphics[width=0.45\textwidth]{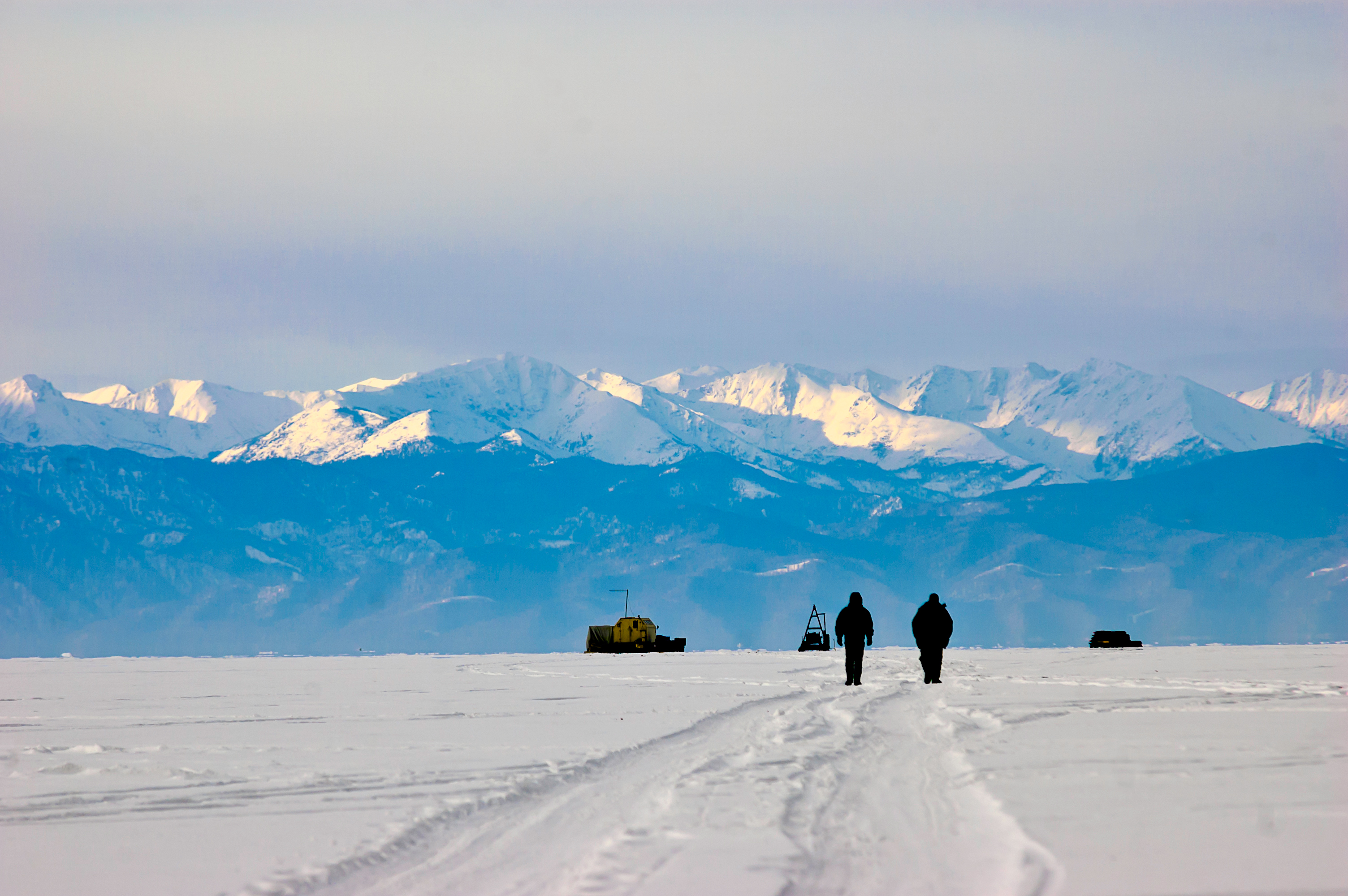}\hspace{2pc}%
    \caption{An overview of the launch site on the Baikal Lake from the shore. The background mountains are more than 50 kilometers away what indicates good atmosphere transparency on ground level.}
\label{fig:baikal_atmo}
\end{center}
\end{figure}

Our own observations of the atmosphere show that during the day the visibility degraded due to high humidity (as is the case shown in Fig.~\ref{fig:baikal_snow}), but in the evening recovered back. In~Fig.~\ref{fig:baikal_atmo} the background mountains are 55--56~km away therefore the horizontal visibility was good. In the nights the mountains were also visible. The Milky Way was clearly visible during all measurement nights. 

Another atmosphere property that was monitored during measurements was its density. The optical transparency is a crucial property for CL-reliant methods of EAS studies since it directly influences the measured light fluxes and later the primary particle energy estimations. However, the atmosphere density profile is also vital for the primary particle type studies since it affects the altitude of shower development and its Cherenkov light lateral distribution function (CL LDF). The influence of the selected atmosphere on the CL LDF and its properties are discussed below in section~\ref{sect:atmosphere-profile}.

The atmosphere density profile was reconstructed basing on air pressure and temperature measurements during the initial climb in the beginning of the night and during the descent at the end. During almost all measurement nights the atmosphere remained stable. Stable atmosphere conditions were expected due to the known Siberian High phenomena. However, during two flights 2012-3 and 2013-5 a rapid change in the atmosphere profile was observed (see Fig.~\ref{fig:density}) followed by rapid weather worsening (clear sky changed to cloudy, wind became stronger with gusts), which, unfortunately, is indicative of the climate change and observed weakening of the Siberian anticyclone. 



The measurements were performed in the end of winter season when the Baikal lake was covered with thick ice. The snow coverage of the ice differs from year to year and in different areas of the lake according to prevailing winds. However in winters 2010--2013 the lake area near the launch site was covered with a thick (up to 40~cm) layer of snow with occasional snowfalls. In Fig.~\ref{fig:baikal_snow} the snow coverage in 2012 is shown. This was the typical coverage during all measurements. The snow reflection properties were controlled using a photometer. The influence of the snow state on its reflecting properties has been discussed in detail in our article on the simulation of the SPHERE-2 detector~\cite{Ant19}.

\subsection{Detector orientation and telemetry data}
\label{sect:telemetrydata}

\begin{figure}[tb]
    \includegraphics[width=0.45\textwidth]{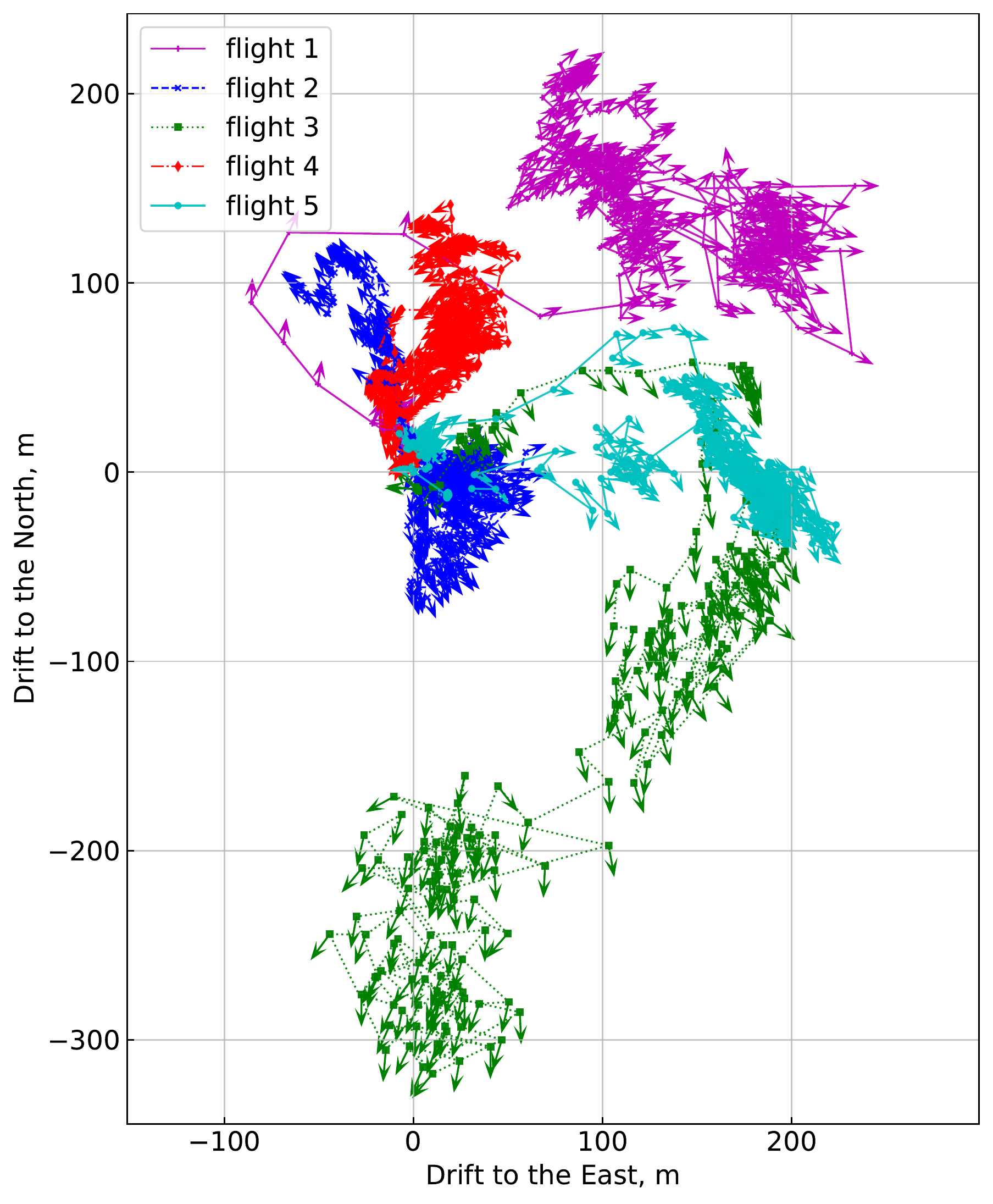}\hspace{2pc}%
    \caption{The SPHERE-2 detector drift in 2013. The arrows indicate the detector magnetometer orientation. The start point is located at zero coordinates.}
\label{fig:gps_compass}
\end{figure}

\begin{figure*}[tb]
    \begin{minipage}[t]{0.48\textwidth}
    \centering
       \includegraphics[width=\textwidth]{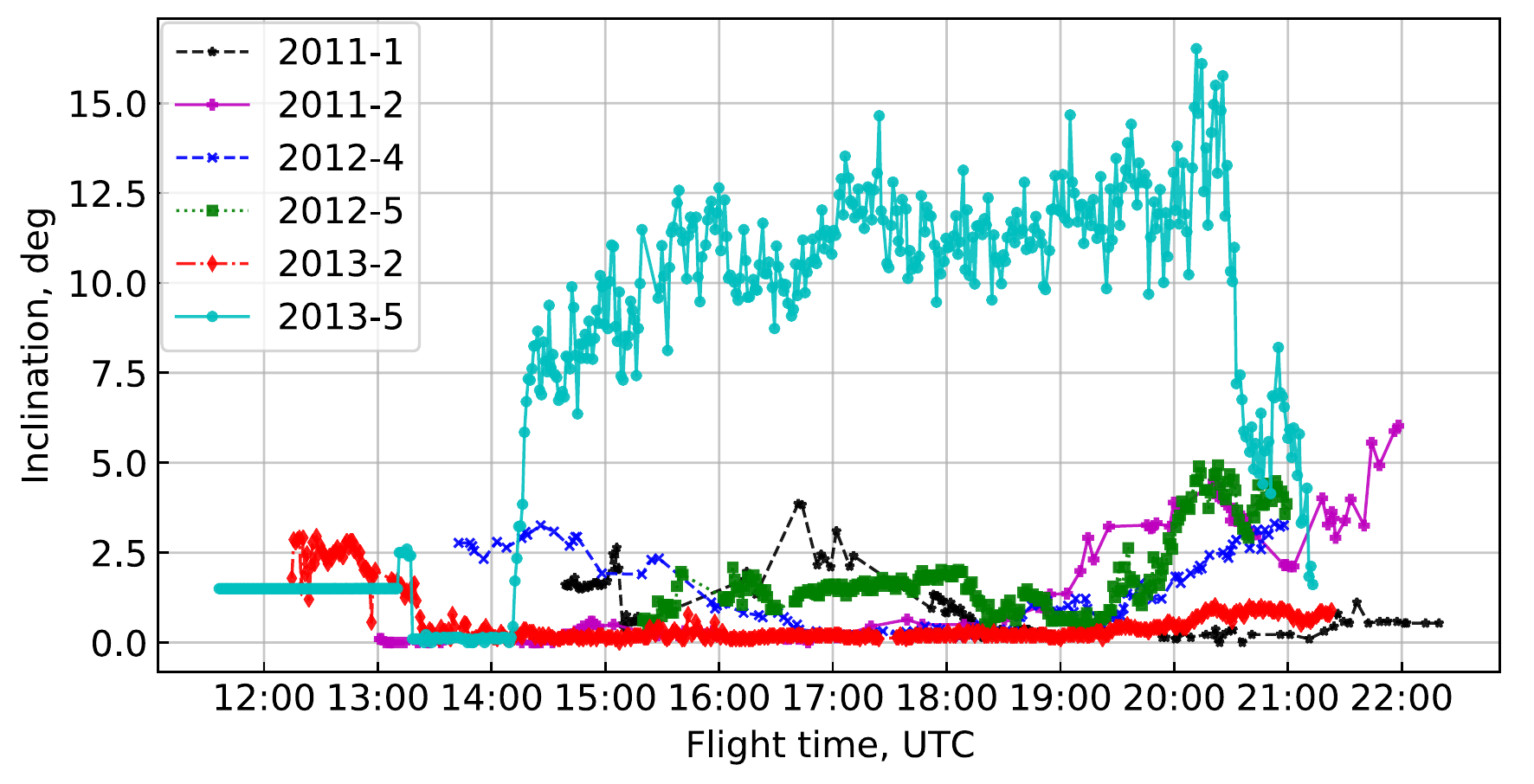}
       \caption{The detector inclination according to the inclinometer sensor during several flights in 2011-2013.}
    \label{fig:inclination} 
    \end{minipage}
    \hfill
    \begin{minipage}[t]{0.48\textwidth}
    \centering
       \includegraphics[width=\textwidth]{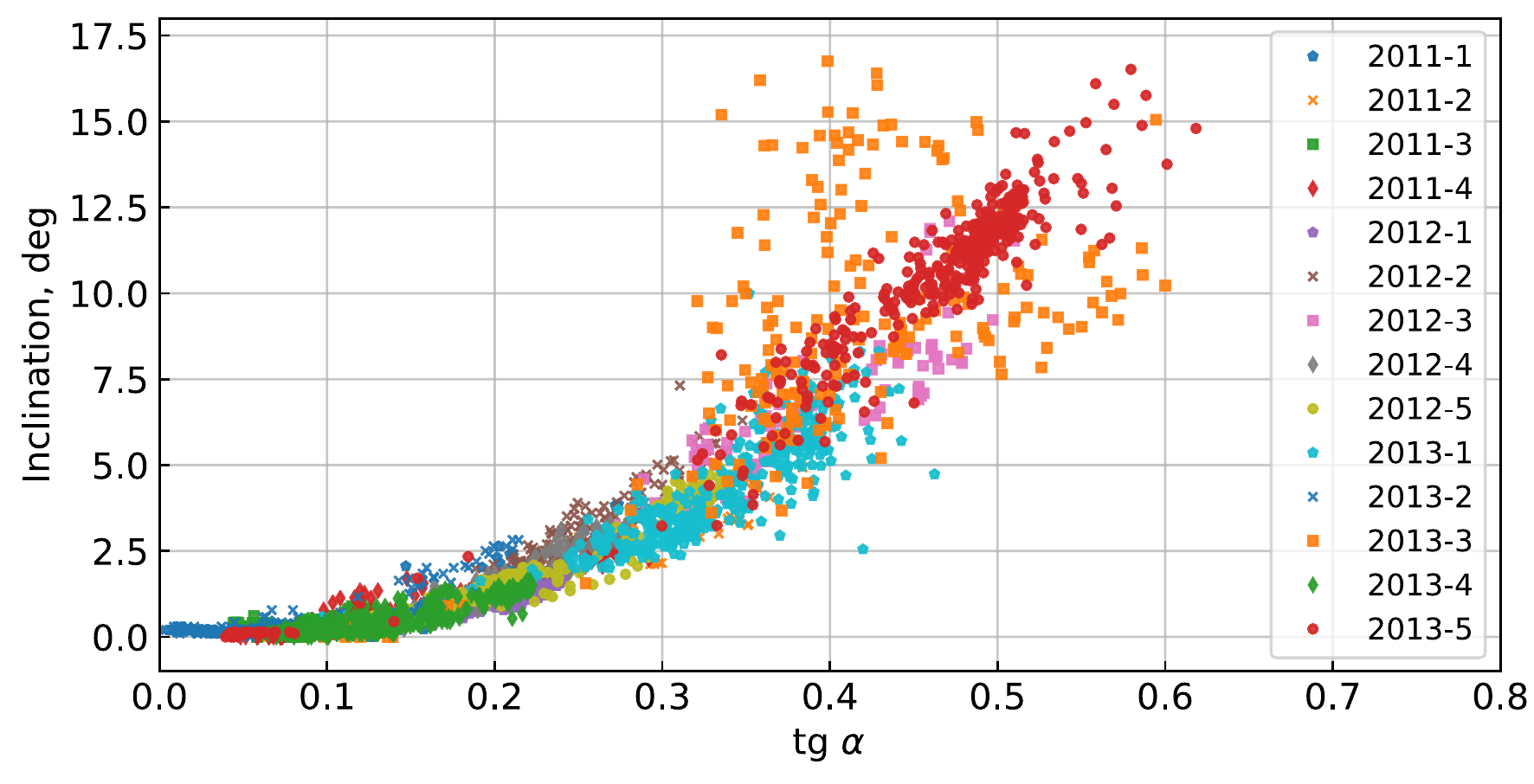}
       \caption{Detector inclination against the detector drift from the start point to the detector altitude ratio which roughly translates into the tether inclination angle. See text for details on detector behavior.}
\label{fig:drift-inclination}
   
    \end{minipage}
\end{figure*}

The \mbox{SPHERE-2} detector's position and inclination depend on wind conditions near the detector, see. Fig.~\ref{fig:gps_compass} and Fig.~\ref{fig:inclination}. In Fig.~\ref{fig:gps_compass} the position of the detector with the magnetometer data is shown for some flights of the 2011--2013 seasons. The position was reconstructed using GPS data with the launch point being located at zero coordinates. The detector orientation measured by the magnetometer is indicated by arrows. Figure~\ref{fig:gps_compass} demonstrates that wind was quite constant during most of the time and the balloon position was rather stable and varied slowly.  

However, due to the construction of the suspension system to the balloon the detector was deflected by the wind to a certain degree in the corresponding direction away from the launchpad. In Fig.~\ref{fig:inclination} the inclination of the detector is shown for some flights of the 2011--2013 runs. During some of the flights the detector was in a nearly vertical position (2011-1 or 2013-2), while in the others (2013-5) it was tilted to a significant angle. But this declination was steady and varied slowly over time. The maximum inclination angle recorded was about 20$^\circ$. In Fig.~\ref{fig:drift-inclination} the dependence of the detector inclination angle from the ratio of the drift distance from the starting point to the flight altitude (roughly the tangent of the balloon tether angle) is shown for all experimental flights. This ratio, as expected, indicates wind strength which, in turn, defines the inclination angle. Strong fluctuations in flight 2013-3 (orange squares) occurred near midnight and coincided with a rapid change in the atmosphere profile and weather worsening, what resulted in early detector landing due to bad flight conditions. The detector inclination had a significant impact on the experiment `geometry' and was taken into account at the stage of EAS parameters reconstruction.

\begin{figure*}[thb]    
    \begin{minipage}[t]{0.48\textwidth}
    \centering
 \includegraphics[width=\textwidth]{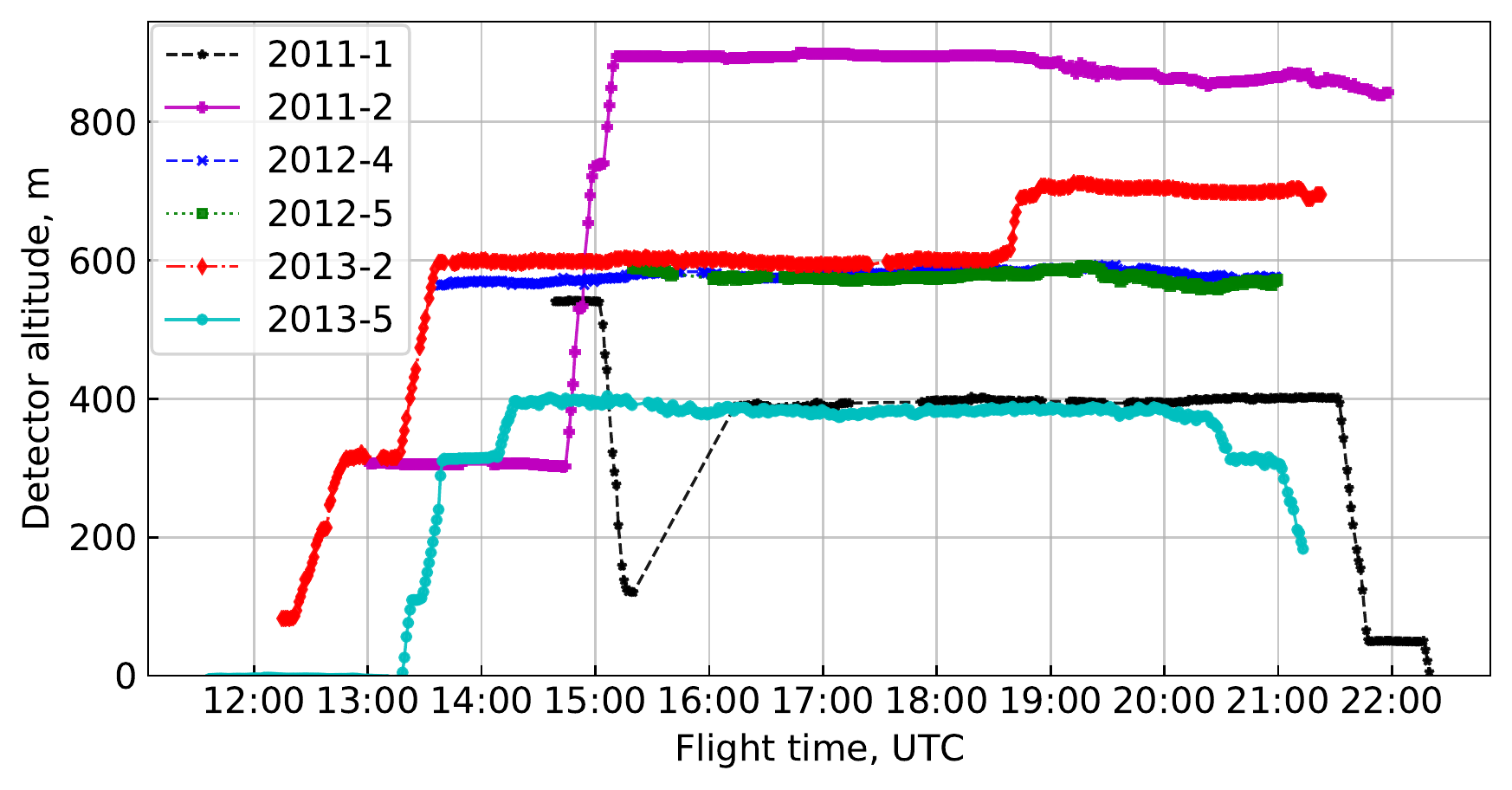}
    \caption{Altitude of the SPHERE-2 detector carried by the BAPA tethered balloon according to the GPS module data during 2011--2013 flights.}
    \label{fig:height}
    
    \end{minipage}
    \hfill
    \begin{minipage}[t]{0.48\textwidth}
    \centering
\includegraphics[width=\textwidth]{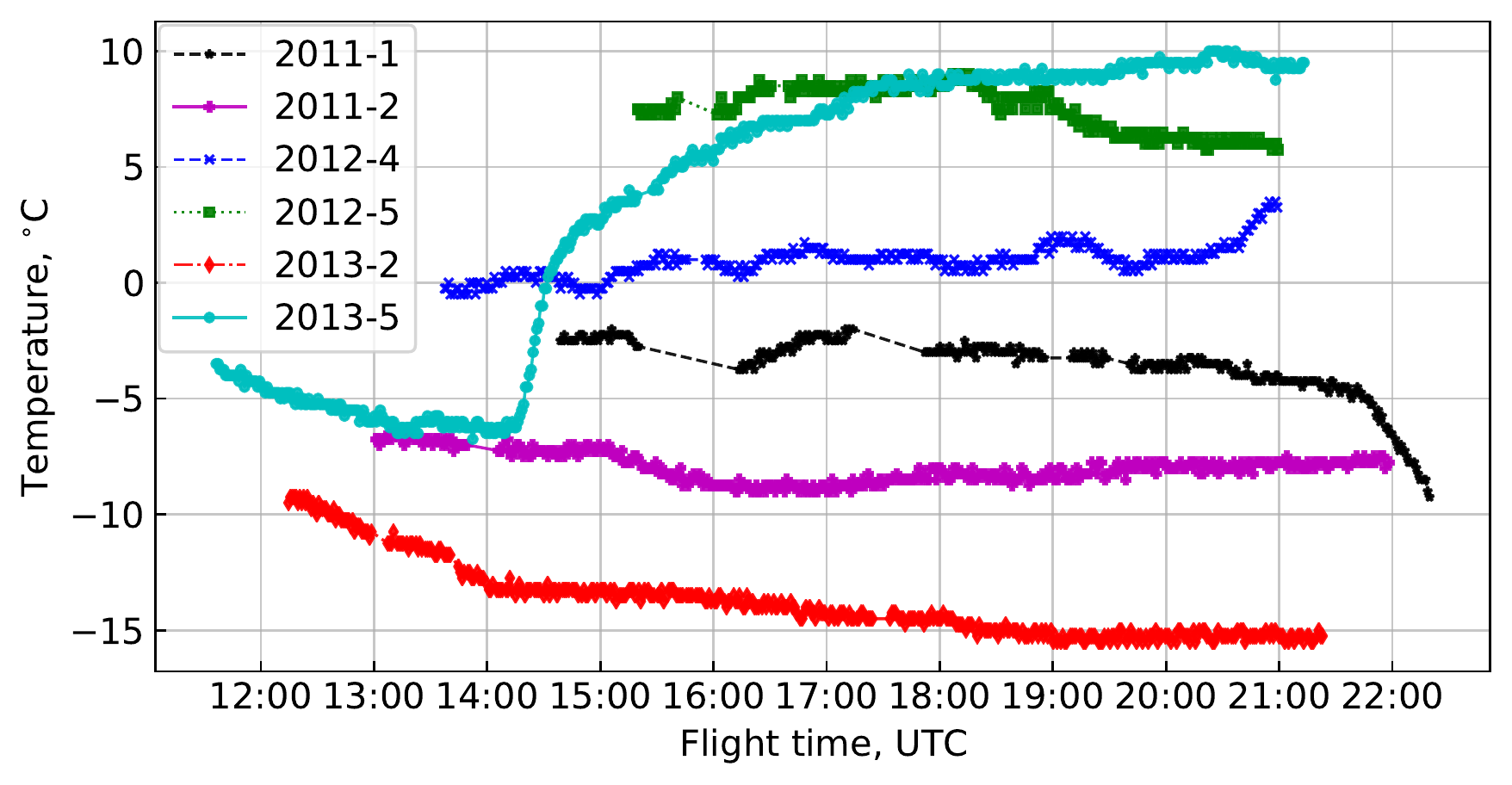}
    \caption{Air temperature near the PMT mosaic during 2011--2013 runs.}
    \label{fig:temperature}
    
    \end{minipage}
\end{figure*}

\begin{figure}[tb]
    \includegraphics[width=.48\textwidth]{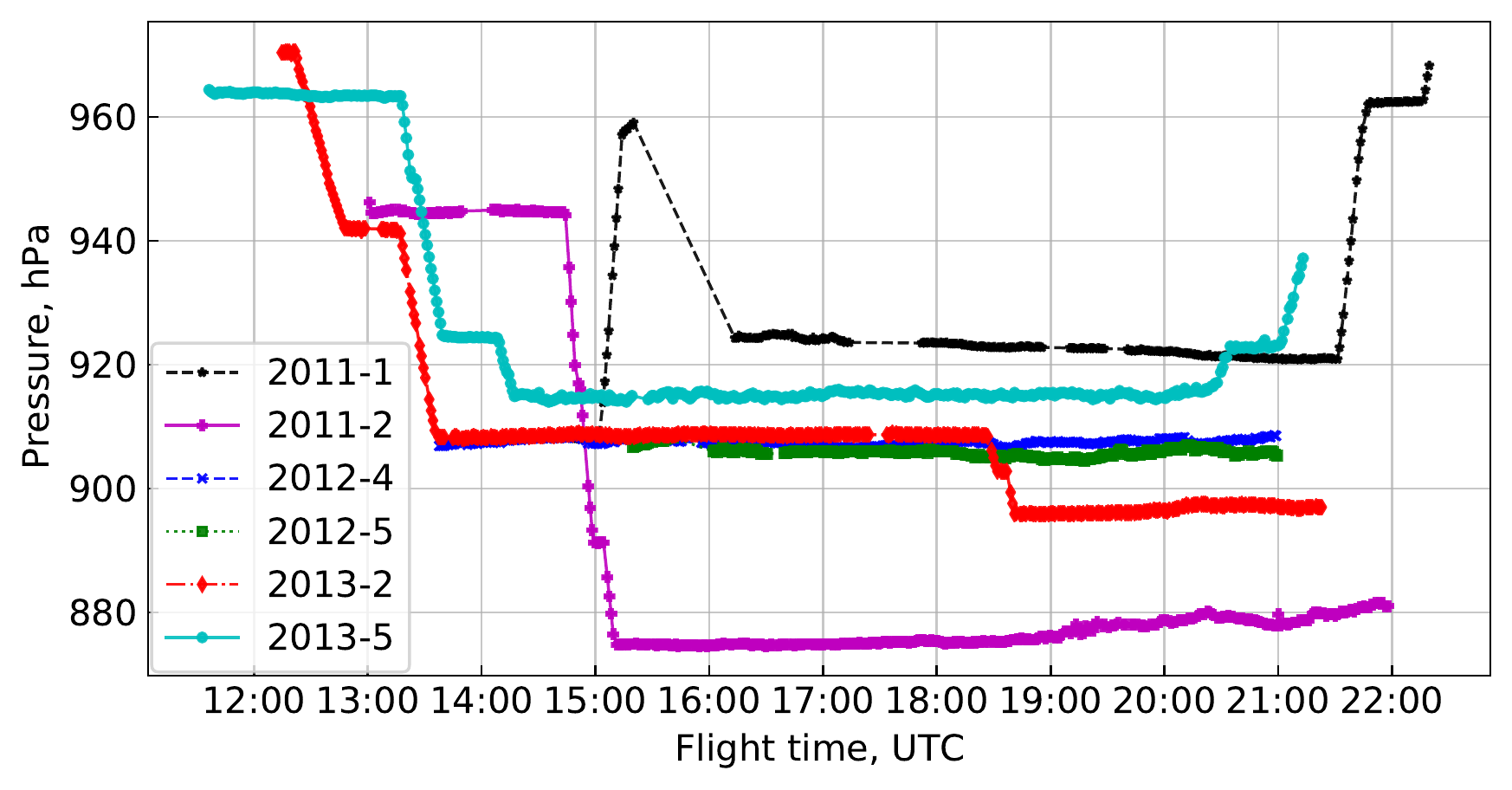}
    \caption{Air pressure according to the barometer sensor data during 2011--2013 flights.}
    \label{fig:pressure}
\end{figure}

Detector altitude above the lake surface according to the GPS data is shown in Fig.~\ref{fig:height}. The GPS module had some specific properties that resulted in several minor corrections in the altitude (detailed description see below in section~\ref{sect:gps_correction}). The surrounding air temperature and pressure that were monitored for the control of atmospheric properties are presented in Fig.~\ref{fig:temperature} and Fig.~\ref{fig:pressure} respectively. Detailed description of the real atmosphere profile and overview of its impact on the data analysis is given below in section~\ref{sect:atmosphere-profile}.

\subsection{PMT currents variations}

\begin{figure}[tb]
    \includegraphics[width=0.48\textwidth]{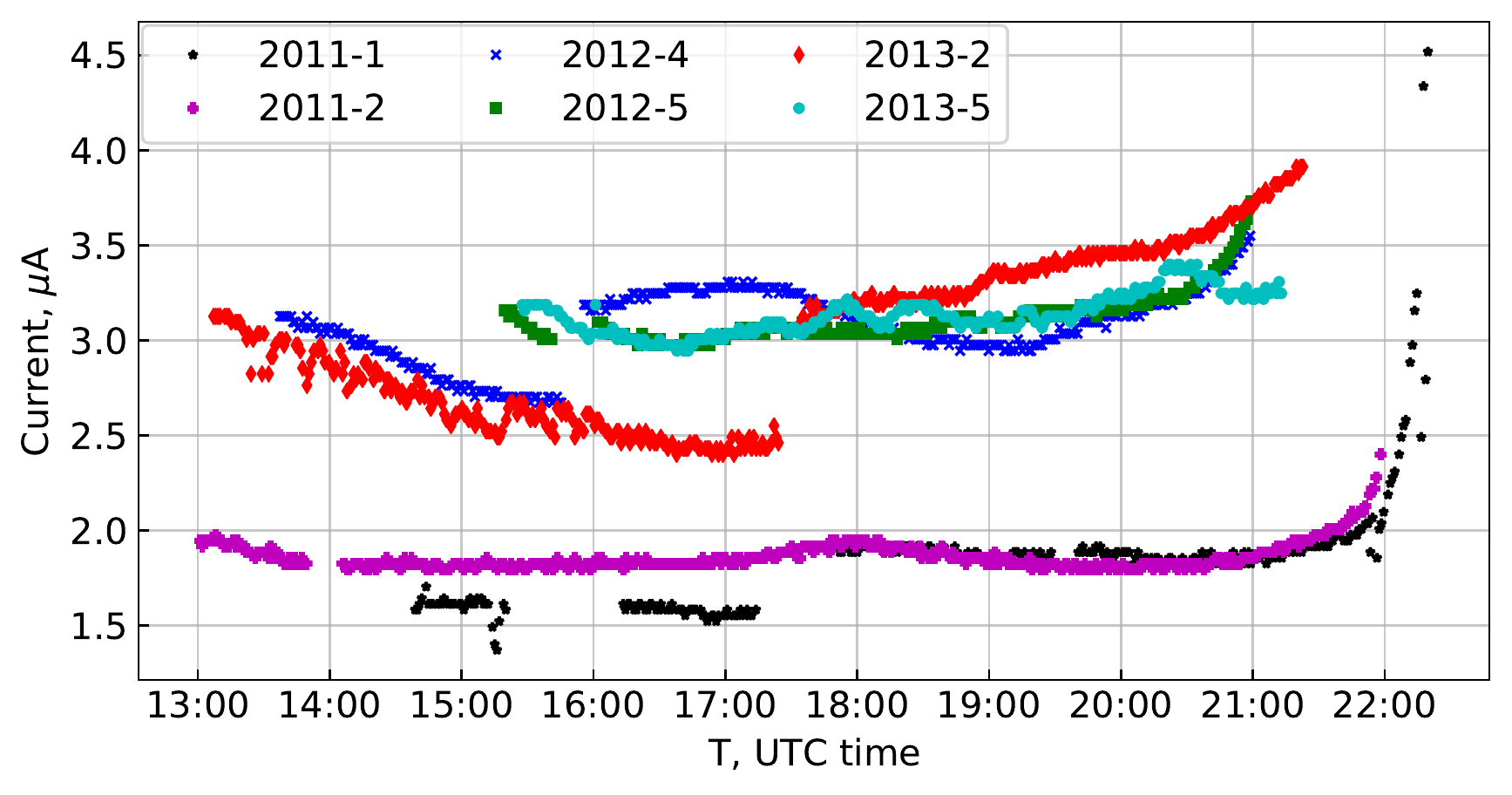}
    \caption{The central PMT currents in different flights.}
\label{fig:current}
\end{figure}

 Fig.\ref{fig:current} shows the anode current of the detector mosaic central PMT for some of the flights. In 2008--2011 seasons in the central position of the mosaic a FEU 84-3 PMT was used and in the seasons 2012--2013 --- a Hamamatsu R3886 PMT. The latter has a larger photocathode and amplification. Also for each flight the voltage on the PMTs was set independently and in some cases was changed mid-flight. The current increase during the last minutes of 2011 flights was due to an increase in the illumination at sunrise. In subsequent seasons the measurements were conducted at earlier dates so the sun rose later and did not affect the measurements. But in general, the PMTs currents variations during the observations followed the variation of the background illumination of the snow.

\begin{figure}[tb]
    \includegraphics[width=0.48\textwidth]{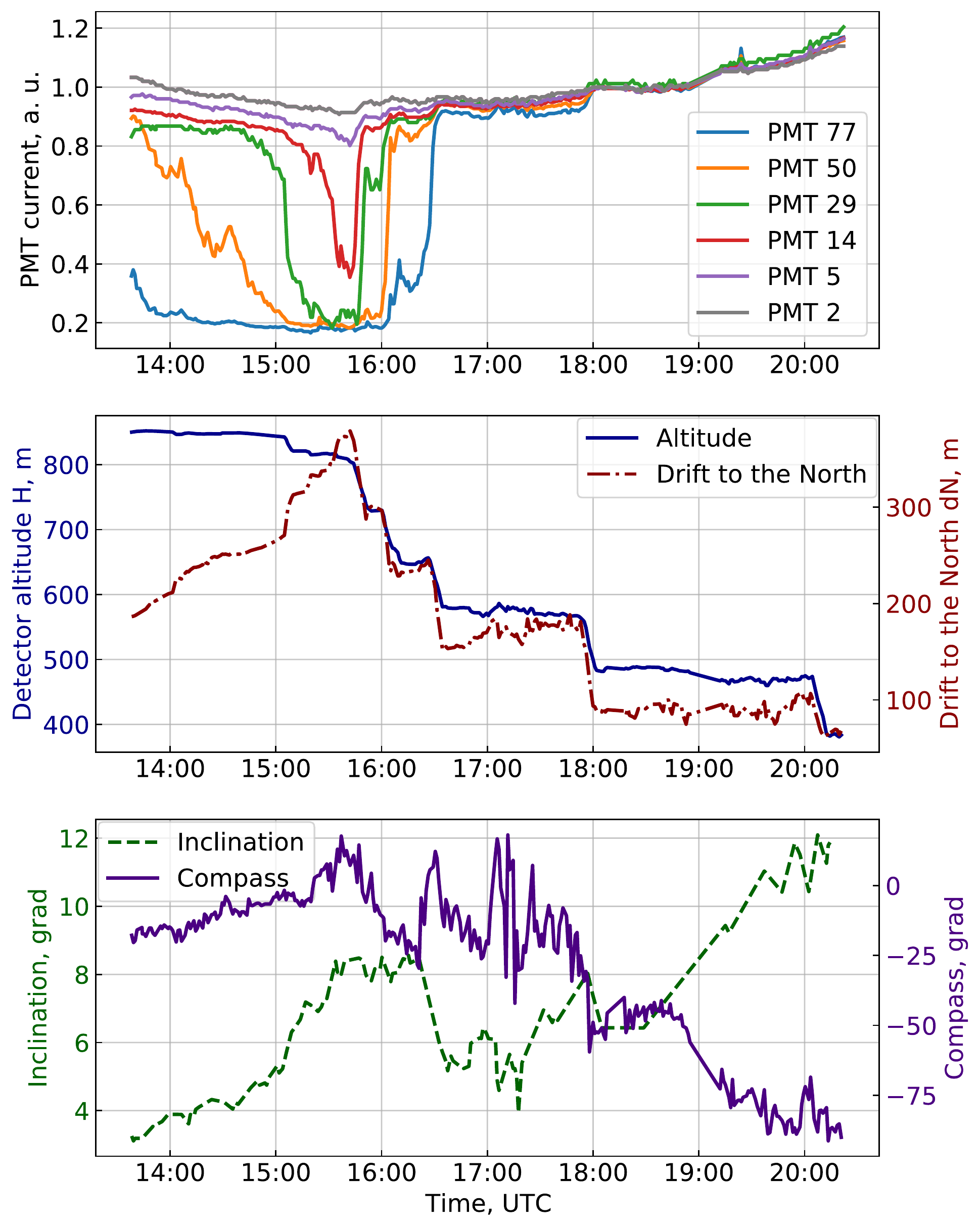}
    \caption{PMT currents, detector altitude, drift to the North, compass and inclination during flight 2012-3.}
    \label{fig:2012-3_currents}
\end{figure}

\begin{figure}[tb]
    \includegraphics[width=0.48\textwidth]{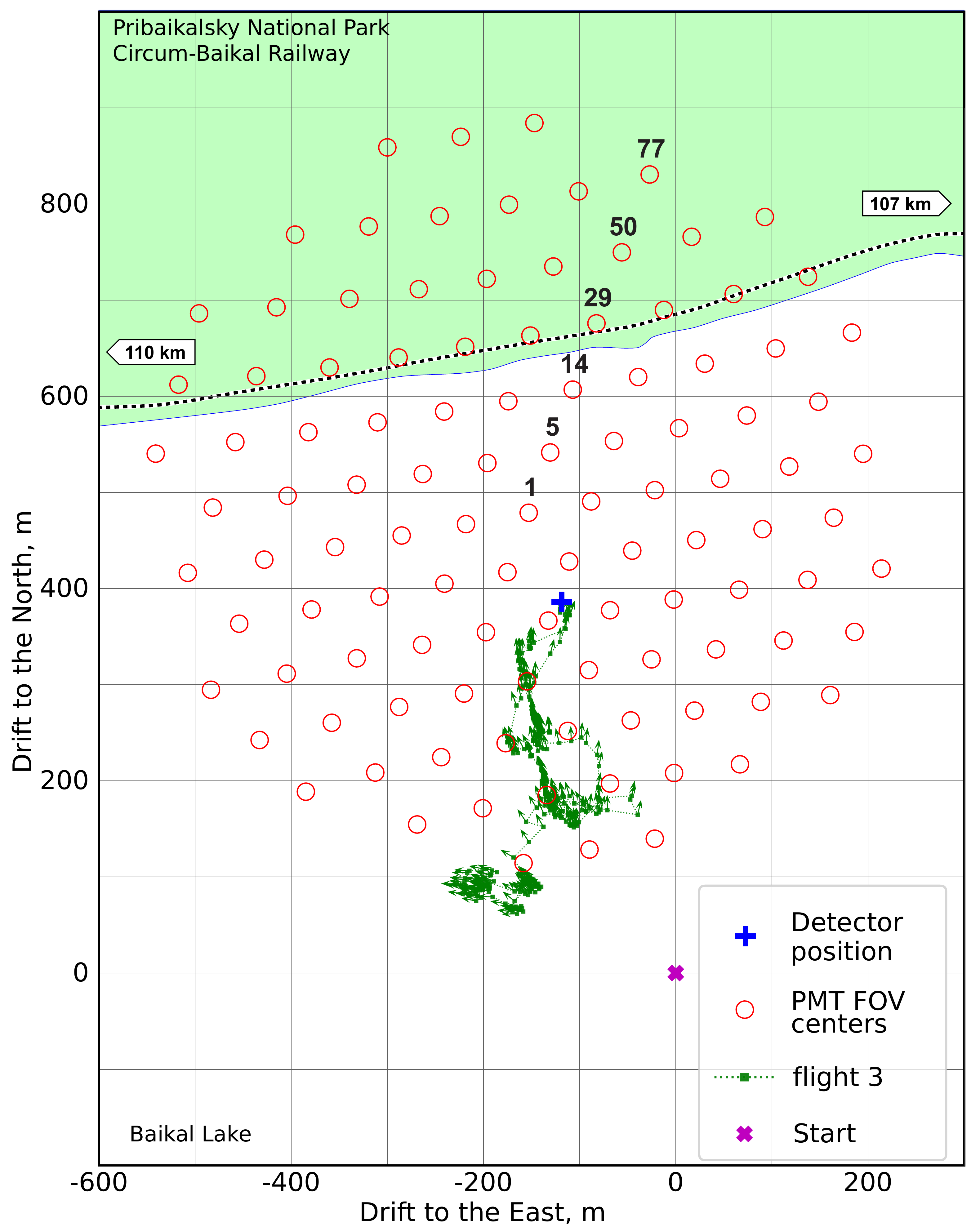}
    \caption{The SPHERE-2 detector drift in 2012-3 flight. Mosaic projection to the show surface is given for the time 15:47 UTC. Detector GPS position at that moment is indicated by the cross.}
    \label{fig:2012-drift}
\end{figure}

\begin{figure}[tb]
    \includegraphics[width=0.48\textwidth]{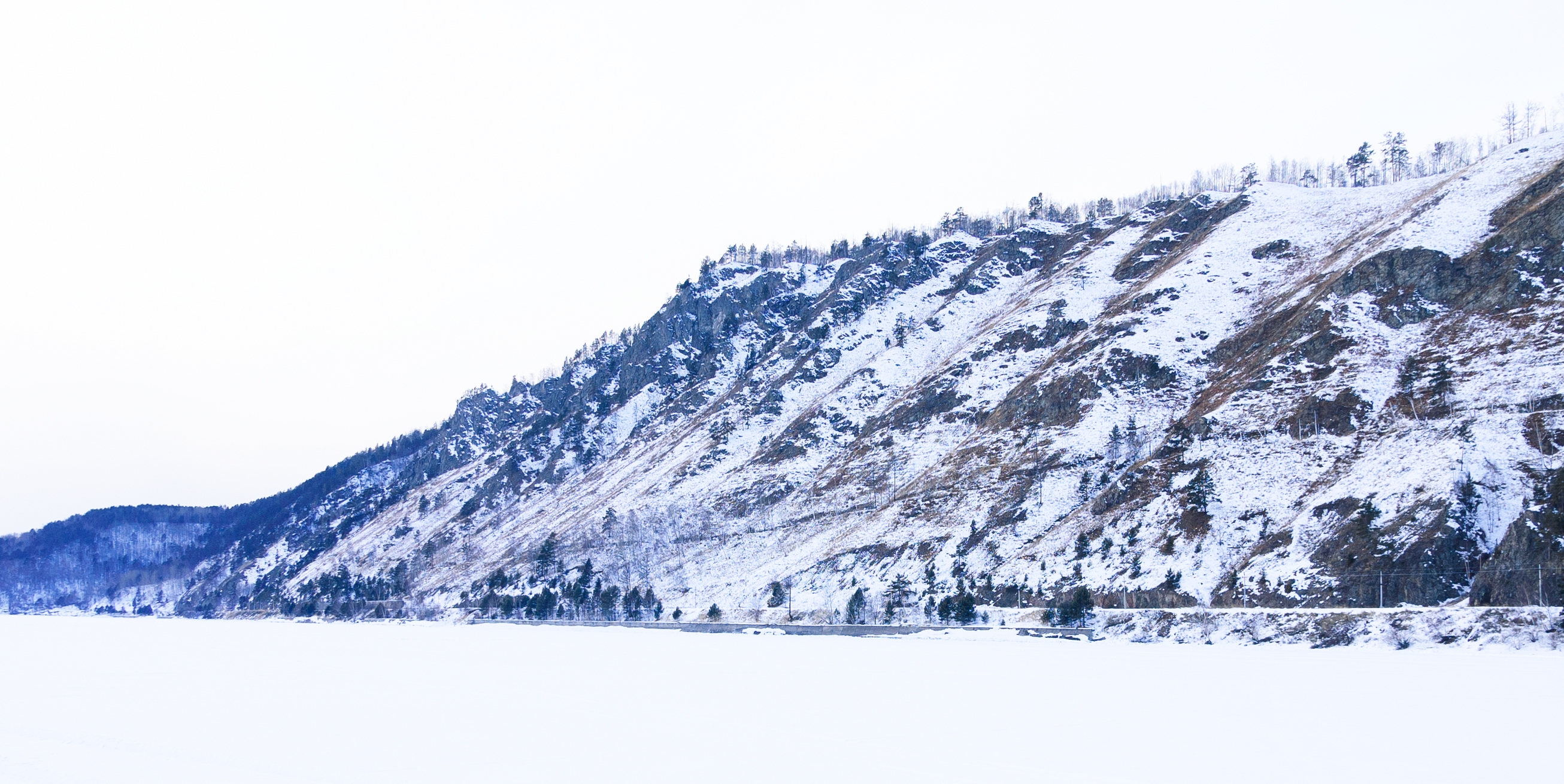}
    \caption{View on the shoreline section `observed' by the detector from the position indicated on Fig.~\ref{fig:2012-drift} is in the middle of the photo.}
    \label{fig:2012--shore-view}
\end{figure}

\begin{figure}[tb]
\centering
    \includegraphics[width=0.35\textwidth]{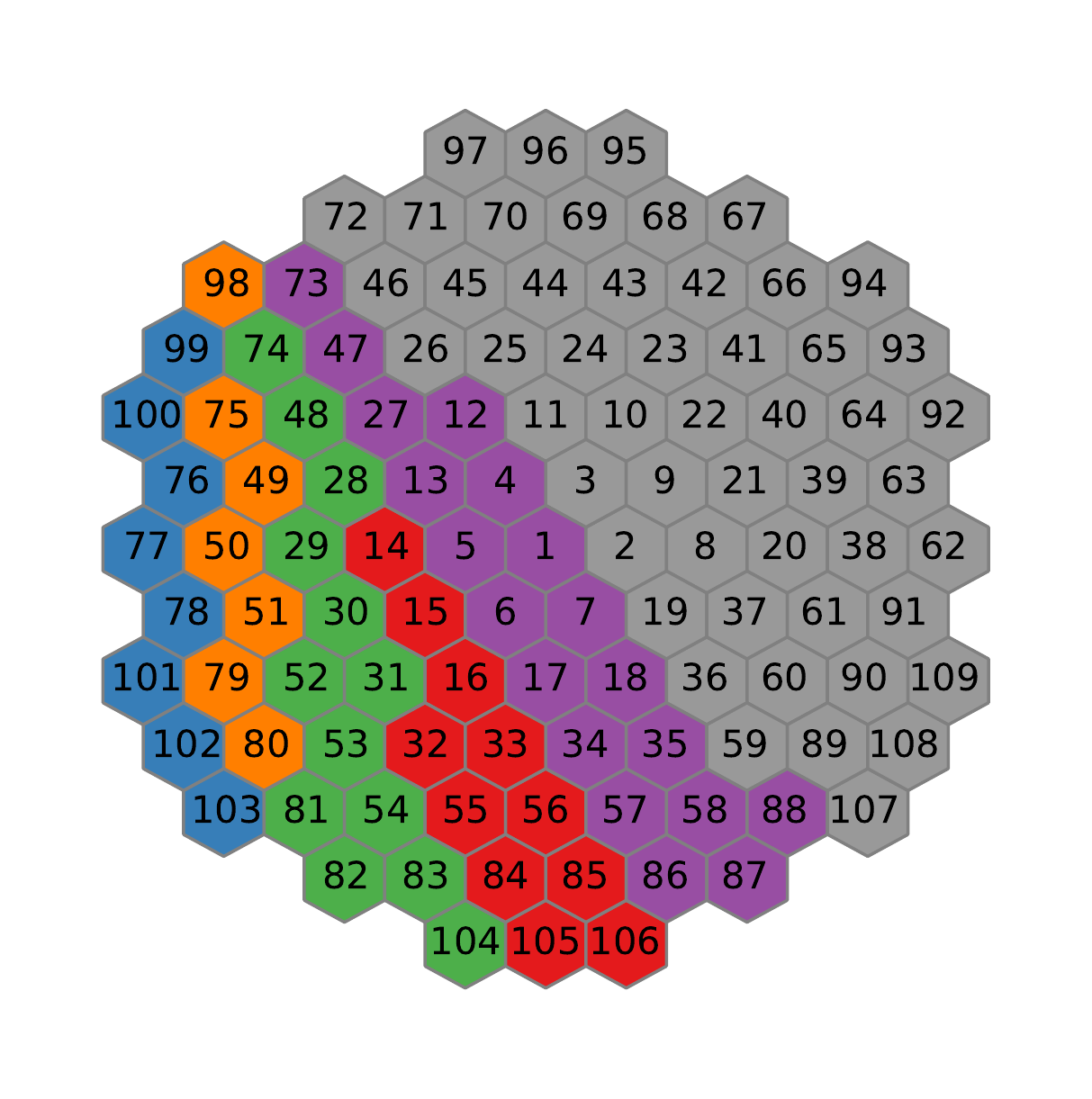}
    \caption{The SPHERE-2 PMT mosaic scheme. PMT numbers are indicated. PMTs are colored according to their current curve type in the 2012-3 flight with colors as in Fig.~\ref{fig:2012-3_currents}.}
    \label{fig:2012-3_shore_image}
\end{figure}

However, in third flight of the 2012 season in part of the PMTs an abnormal drop in the currents was observed.
The registered currents from some of the PMTs (numbers 5, 14, 29, 50 and 77) are shown in Fig.~\ref{fig:2012-3_currents}, top panel. These PMTs form a straight line on the PMT mosaic and the currents dropped and restored in them in the same order as they are positioned. This behaviour was observed for a number of PMTs with current drops occurring at the same time. In Fig.~\ref{fig:2012-3_shore_image} the PMTs are colored according to similar patterns in the current drop and restore times.

This anomaly was observed in the first half of the flight during observations made at an altitude of 850~m. At that time according to the GPS the balloon was drifting north from the starting point. Detector altitude and drift are shown in~Fig.~\ref{fig:2012-3_currents}, central panel. Due to wind strengthening at that altitude the balloon was subsequently lowered to 480~m and the currents returned to their normal behaviour. 

Analysis of the detector position and tilt showed that at that time it was observing a part of the shoreline and land surface. At this location the shoreline is a narrow artificial ledge built for the railroad and further changes into a steep snow-less cliff wall with trees atop (see Fig.~\ref{fig:2012--shore-view}). The reflective properties of both rock and pine trees is way lower than that of snow. In Fig.~\ref{fig:2012-drift} the detector trajectory (green line) and starting point (purple cross) are given. The detector position (blue cross) and corresponding PMT centers projections onto the surface are given for time point 15:47 UTC with the detector inclination of 8$^\circ$ to the north, the moment of the highest current drop and furthest drift towards the shore.The behavior of the currents in each PMT is consistent with its crossing of the shoreline. The PMTs that show no drop in their currents by our estimations always observe a clear snow surface of the lake. In all other flights no instances of detector drift to the shore was observed. 

This observation itself confirms the consistency of the detector orientation and position monitoring.

\section{Telemetry analysis}

The position and inclination of the SPHERE-2 detector are important values for the correct reconstruction of the CL characteristics of registered EAS. In contrast to Cherenkov ground-based arrays, such as~\cite{Yakutsk19,TUNKA133}, the SPHERE-2 detector has a variable registration area depending on both altitude and inclination.

\subsection{GPS altitude correction with barometer data}
\label{sect:gps_correction}

\begin{figure}[tb]
    \includegraphics[width=0.48\textwidth]{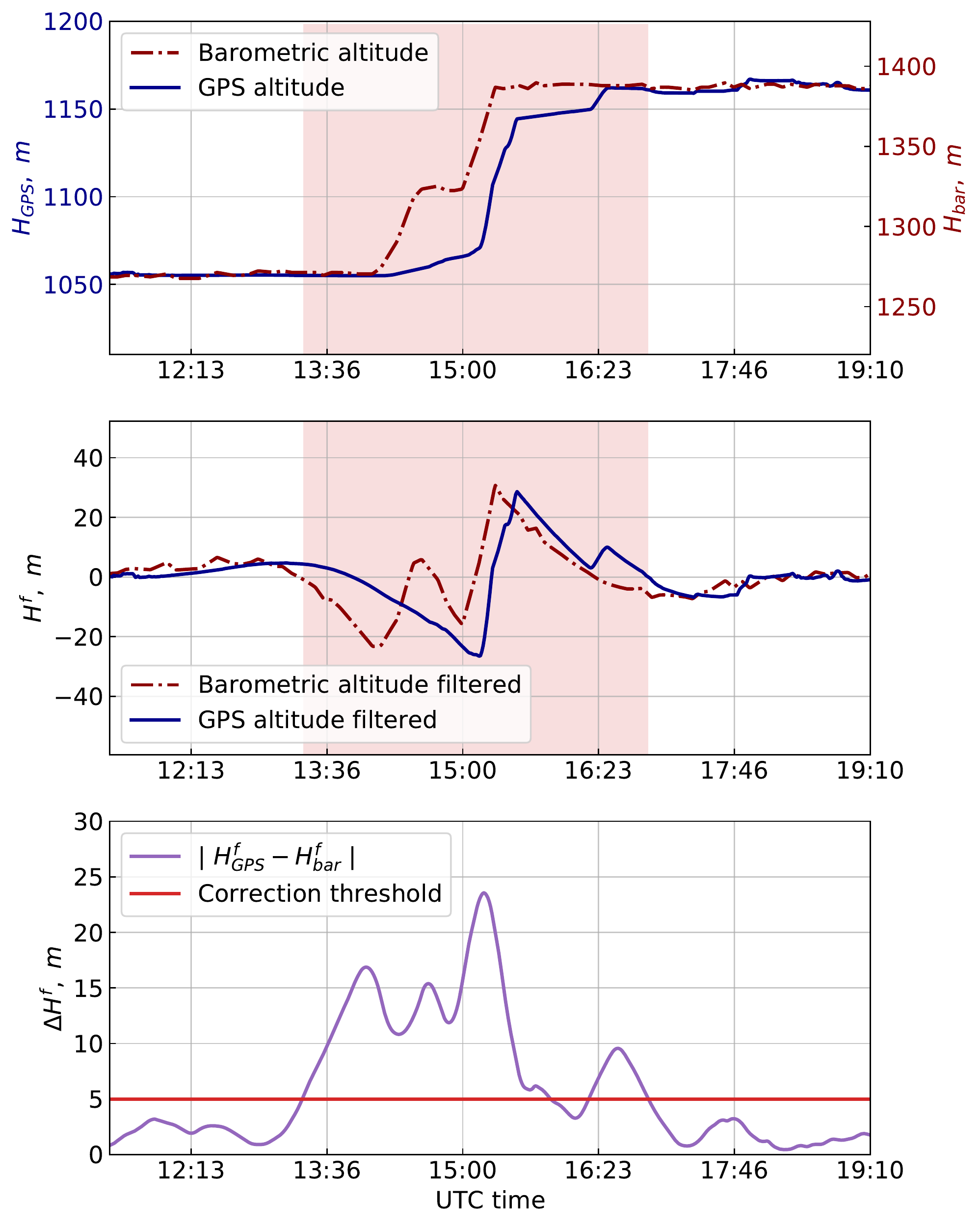} 
    \caption{An example of the anomalous GPS altitude behavior, as compared with the `barometric' altitude (top panel) during flight 2013-2. Same data with a high-pass digital Butterworth filter applied (middle panel). Smoothed absolute delta of high-passed altitudes from the middle panel, used as a gate with the threshold set by the red line; intervals above this line correspond to the red bands on the other panels, which are the intervals subjected to GPS data correction.}
\label{fig:h_corr}
\end{figure}

Thorough telemetry data analysis revealed that at several periods of time the recorded altitude from the GPS data was inadequate. Namely, after rapid changes in altitude due to the balloon ascent or descent the GPS registered height lagged behind from the corresponding pressure change and varied in a smoothed fashion. The example of such behavior is shown in Fig.~\ref{fig:h_corr}, top panel. We were unable to find out the exact source of this distortion, but our best guess is that it's the result of the GPS module's internal error correction algorithm. The module's primary application is for naval/aerial navigation~\cite{GPS-module-specs} and it probably may interpret sudden changes in altitude without a proper horizontal speed as an error.

Affected time intervals are relatively short for most flights, but some events were recorded during these times. In order to correctly estimate the detector altitude for those events, and to maintain the overall data consistency we attempted to correct these smoothed GPS altitude measurements using the pressure data from the barometer, which has a strong and well-understood dependence on the altitude. The correction was made for total of 220 minutes of all flights for which the difference between the two altitudes was more than 10~m. During this time 68 triggers were registered.

First of all, in order to identify the affected time intervals we calculated the approximate `barometric height' from the pressure data using the inverted barometric formula $H_{bar} = H_0 - a \log (P/P_0)$ with parameters $H_0 = 456~\textrm{m}$ (Baikal lake surface level), $P_0 = 1000~\textrm{hPa}$ (close to average March pressure on the Baikal surface) and $a = {RT}/{Mg} \approx 8400~\textrm{m}$ (standard value for the barometric formula). Next, we applied a high-pass digital Butterworth filter to both GPS and barometric altitude data to compare only the fast variations of both values. The cutoff frequency for the filter was set to $0.416~\textrm{mHz}$ ($T=40$ min with telemetry data being recorded every $10~\textrm{sec}$) and slope set to around $12~\textrm{db per octave}$. The resulting high frequency components are shown on Fig.~\ref{fig:h_corr}, middle panel.

It is clear that for stable altitude and valid GPS measurements the GPS and barometric heights are subjected to fast, somewhat correlated fluctuations (they are not completely correlated probably due to the GPS error estimated to be of the order of magnitude of approximately a meter). In contrast, intervals of smoothed GPS data show divergence of these two values. We calculated the delta between the filtered GPS and barometric altitudes and applied a moving average with a $6~\textrm{min}$ window. We then identified intervals for correction as those that feature the smoothed delta above a threshold of $5~\textrm{m}$ in absolute value. The threshold was chosen so that  under regular conditions the smoothed delta did not exceed it, as it has a normal-like distribution with a standard deviation of around $3~\textrm{m}$. The smoothed delta and threshold are shown in Fig.~\ref{fig:h_corr}, bottom panel; the correction interval is shown on the top panel by the red band.

After the intervals were picked, for each of them we found two adjacent intervals $2~\textrm{min}$ each with correct GPS data and once again used the barometric formula to estimate the actual altitude inside the interval. In this case it was used to describe the local behavior of $H(P)$. We fitted $(P, H)$ data points from adjacent regions with the same barometric function and replaced the smoothed GPS data with the new `local barometric height'.

\begin{figure}[t]
    \includegraphics[width=19pc]{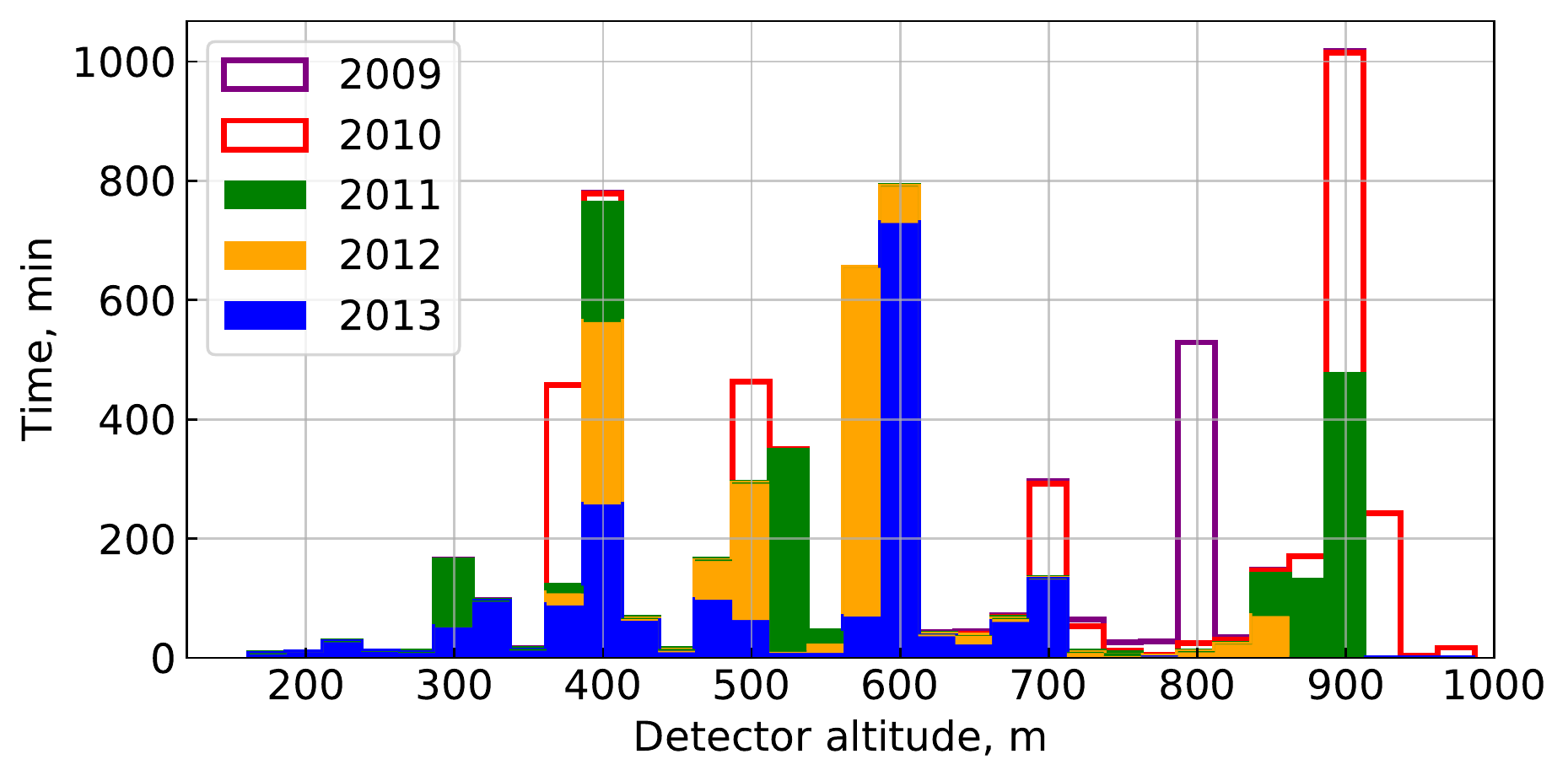}%
    \caption{Altitude distribution over the experiment time.}
    \label{fig:time_on_altitude}
\end{figure}

The total duration of the intervals subjected to corrections was small. The total distribution of the time spent by the detector at each height is presented in Fig.~\ref{fig:time_on_altitude}. Most of the time the detector spent at 400, 500, 580 and 890~m. However, the actual distribution of the registered events across the altitudes is a more complex question since the trigger settings and overall detector performance changed over time and across seasons. Careful analysis of this question will be covered in following publications.

\subsection{Atmosphere density profile}
\label{sect:atmosphere-profile}

After the corrections were applied to GPS altitudes the atmosphere density profiles were reconstructed. Direct measurements with an on-board barometer and thermometer were taken only at altitudes below 900~m above the ice, therefore information about higher atmospheric conditions is not available. To obtain the best extrapolation of this data we had to use a set of parameterized atmospheres provided by the CORSIKA software~\cite{hec98}.

In CORSIKA the atmosphere is described in terms of mass overburden $t(H)$ that is parameterized as a piece-wise continuous function which has exponential behavior in the lower four layers and linear in the highest one. CORSIKA code provides 26 atmosphere models corresponding to measurements taken at different seasons in different locations around the globe. For each model of the atmosphere our experimental data was fully located in the lowest layer.



\begin{figure}[bt]
\centering
\begin{minipage}[t]{0.48\textwidth}
    \includegraphics[width=\textwidth]{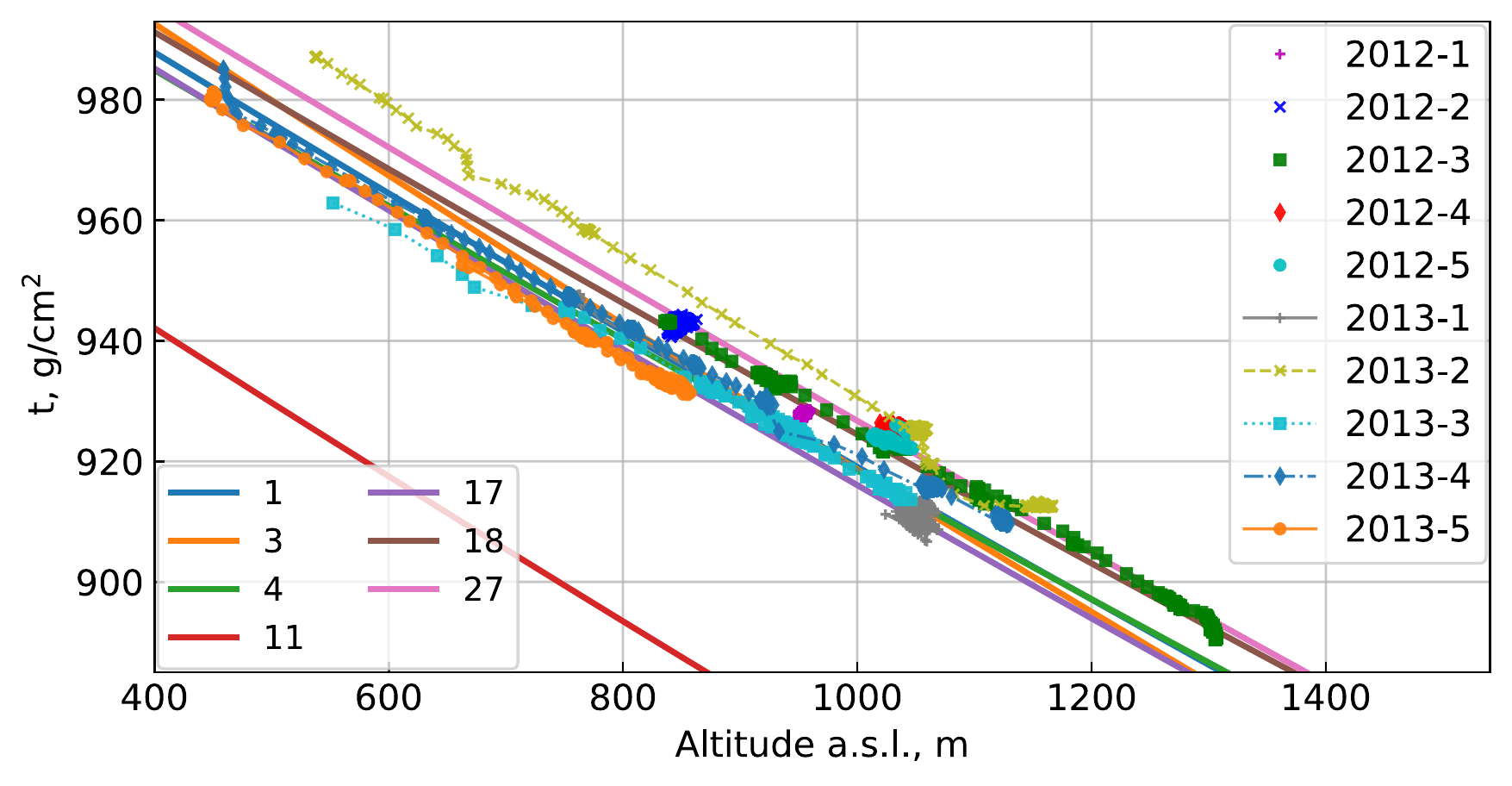}
    \vspace{-1.0pc}
    \caption{Mass overburden versus altitude experimental data (points) in each flight and CORSIKA profiles (solid lines with corresponding model numbers). For preliminary SPHERE-2 modeling and analysis the N0 11 atmosphere was used.}
\label{fig:massoverburden}
\end{minipage}
\vfill
\vspace{1pc}
\begin{minipage}[t]{0.48\textwidth}
    \includegraphics[width=\textwidth]{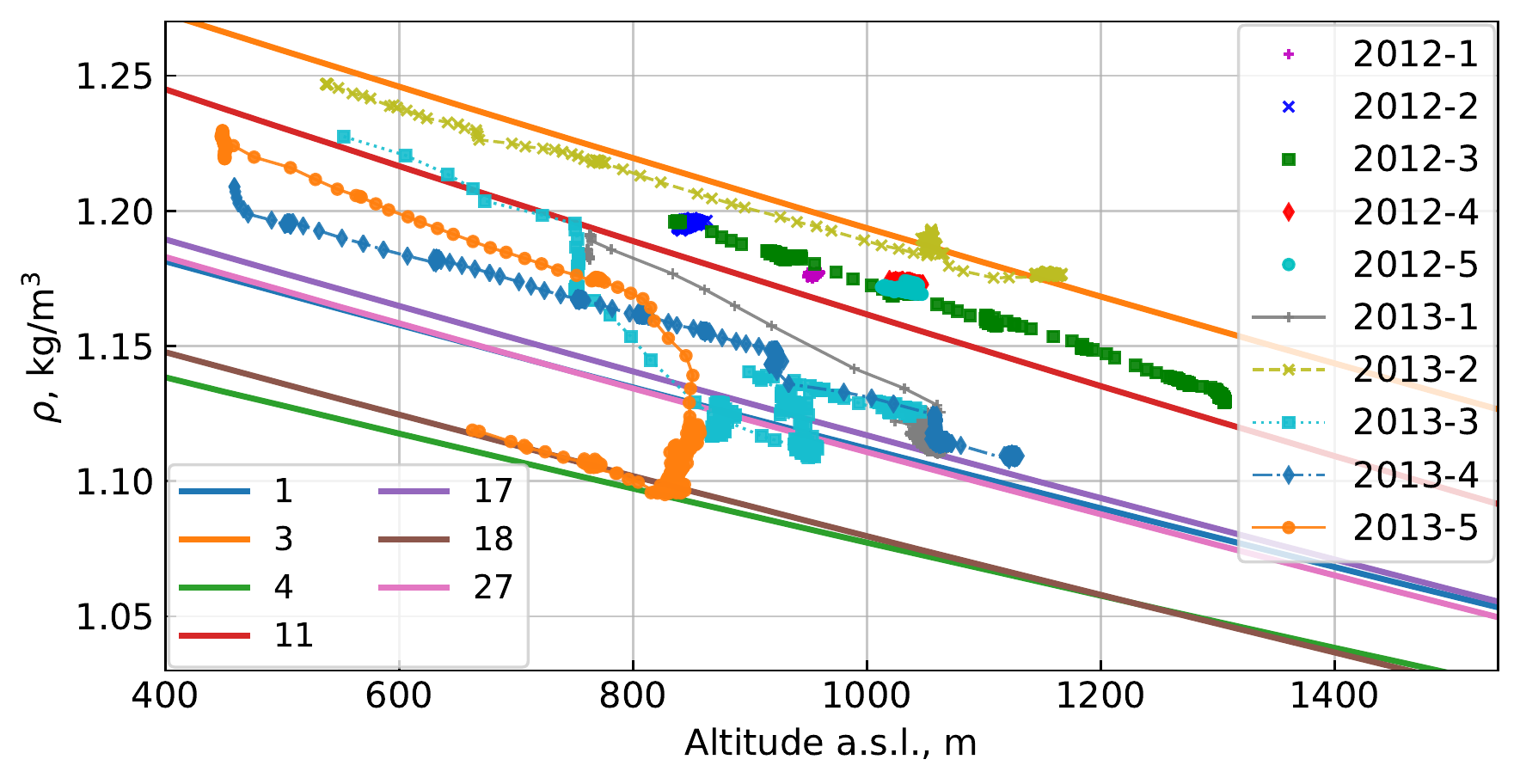}
    \vspace{-1.0pc}
    \caption{Density versus altitude experimental data (points) in each flight and CORSIKA model density profiles (solid lines) same as in Fig. \ref{fig:massoverburden}).}
\label{fig:density}
\end{minipage}
\end{figure}

Mass overburden $t$ data was reconstructed from atmospheric pressure measurements and air density $\rho$ was estimated based on both air pressure and temperature around the balloon. Calculated experimental points for $t$ and $\rho$ for each flight are shown in Fig.~\ref{fig:massoverburden} and Fig.~\ref{fig:density} respectively. CORSIKA atmospheres are shown with solid lines. The profile that was adopted prior to the actual measurements for the preliminary modeling (atmosphere model 11) is shown by the red line. From this comparison it may be seen that the previously picked atmosphere is inconsistent with our experimental data for $t$ in terms of absolute values, but their derivatives lie relatively close. 

\begin{table}[t]
\centering
\caption{Total Cherenkov photons numbers ratios for different CORSIKA atmosphere model pairs: means, variations, relative variations given for 10 PeV primary protons. Zenith angle 15 $\deg$. Sample volume 30 events.}
\label{tab:atmmod}
\vspace{1pc}
\begin{tabular}{|c|c|c|c|}
\hline
model/primary   & mean &  variation   & relative variation \\ 
     pair       &  $m$ & $\sigma$     & $\sigma/m$          \\ 
\hline 
\hline 
 3/4 &  1.015    &  0.0490     &   0.0483   \\
\hline
11/3 &  0.9834    &  0.0511     &   0.0520    \\
\hline
11/4 &  0.9963    &  0.0443     &   0.0445    \\
\hline
\hline
p/Fe &  1.232     &  0.0686     &   0.0557     \\
\hline
\end{tabular}
\end{table}

\begin{figure*}[tb]
 \begin{minipage}[t]{0.48\textwidth}
    \centering
    \includegraphics[width=19pc]{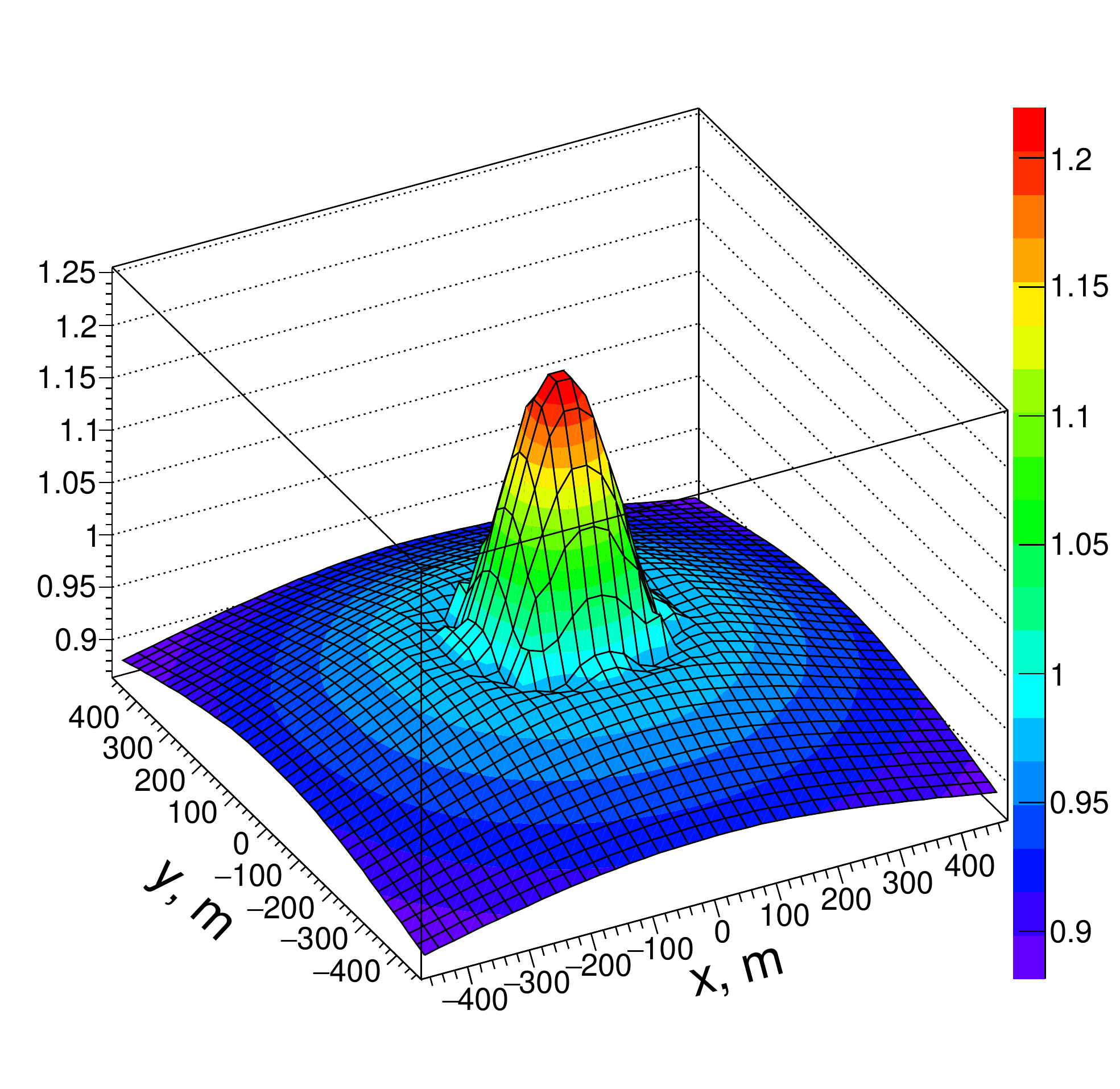}%
    \vspace{-1.0pc}
    \caption{Sample mean of the CL LDF ratio for CORSIKA atmosphere model pair 11/4. Bin size 50~m $\times$ 50~m. 10~PeV primary protons. Zenith angle 15~$\deg$. Sample volume 30~events.}
\label{fig:4d11}
\end{minipage}
\hfill
\begin{minipage}[t]{0.48\textwidth}
    \centering
    \includegraphics[width=19pc]{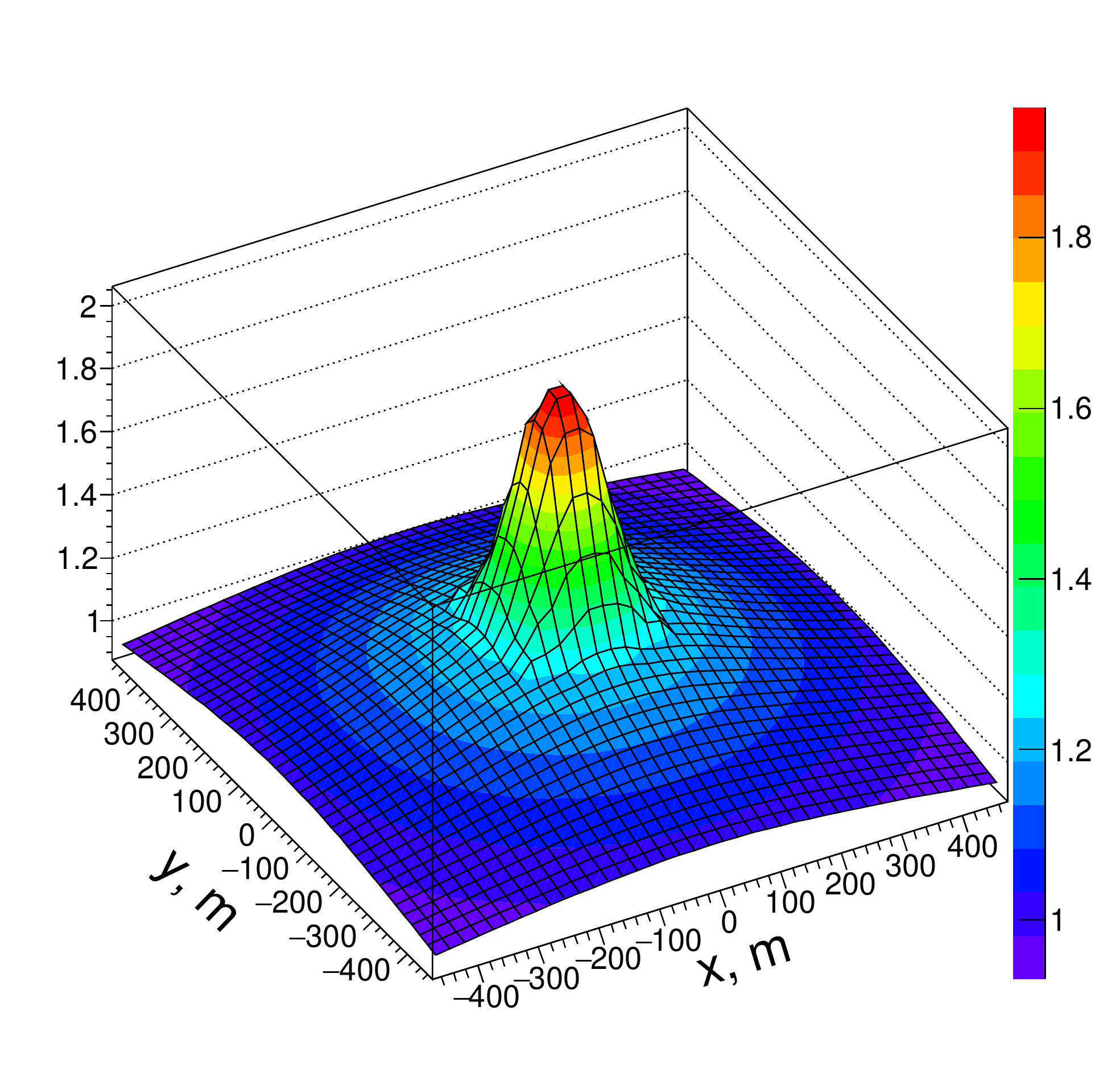}%
    \vspace{-1.0pc}
    \caption{Sample mean of the CL LDF ratio for the nuclear pair proton/Fe for CORSIKA atmosphere model 11. Bin size 50~m $\times$ 50~m. 10~PeV primary protons. Zenith angle 15~$\deg$. Sample volume 30~events.}
\label{fig:pdFe}
\end{minipage}
\end{figure*}

CORSIKA atmosphere model 11 was selected based on available monthly averages from Irkutks airport weather station located 70 km to the North of the balloon launch site. The model was used for preliminary modelling and estimations at the experiment planning stage. However, our data show that this choice was not ideal.

The use of experimental points in $t$ or $\rho$ cannot provide a reasonable choice of the model, but only give a clue as to what the model should be. A full-fledged choice must include data on the vertical profile of the atmosphere up to altitudes of at least 10 km above the lake because Cherenkov light mostly comes from this layer. Still some conclusions on the atmosphere model effect on primary energy and mass estimates can be made by comparing the artificial showers simulated for different models.

Artificial showers initiated by 10~PeV protons were modeled using CORSIKA for atmosphere model 11 resembling the experimental $\rho$ trajectories, as well as for models 3 and 4 roughly following the data on $t$. The results are expressed as ratios of full numbers of Cherenkov photons and CL LDFs. We use primary proton showers here because they reveal the most pronounced atmosphere model effect. 

The lateral distributions were calculated within a vast (3.2~km~$\times$~3.2~km) carpet tiled with 2.5~m~$\times$~2.5~m squares but for the purpose of comparison were smoothed by integrating over 50~m~$\times$~50~m squares approximately imitating the sensitivity spots of the telescope pixels.

The sample volume for each model was 30 showers. Table~\ref{tab:atmmod} shows the mean values and variations of Cherenkov photon number ratios for three pairs of CORSIKA atmosphere models (3/4, 3/11 and 4/11). For the reference in the last row the ratio is given for the same atmosphere model (4) but for different primaries (proton to iron). The ratio of sample mean lateral distributions for pair 11/4 is shown in Fig.~\ref{fig:4d11}. For comparison in Fig.~\ref{fig:pdFe} the ratio is of CL LDFs from proton induced showers to those from iron is shown is the same atmosphere. 


Table data clearly states that substantial changes of the atmosphere model affect the total number of Cherenkov photons and thus the primary energy estimates not more than by 5\% on average. Conclusions on primary mass estimates are not as clear, but the lateral distribution ratio plots indicate the changes in CL LDF to be about 6 to 12\% (12 to 20\%) near the shower core, closer than 80 m, and about -2 to +6\% (0 to +12\%) in the 80--150m circle for the 3/11 (4/11) pair, which definitely affects our mass-sensitive criterion. Our preliminary studies indicate that mass-sensitive parameter based on CL LDF steepness is extremely sensitive to the atmosphere model. In case of the parameter used in our study the value of the parameter vary greater with atmosphere than with primary particle mass. The good knowledge of the atmosphere parameters is crucial for the reliable primary particle mass estimation.

The detailed description of our primary particle mass estimation procedure with accurate evaluation of the atmospheric effects and related uncertainties will be given elsewhere.

\section{Conclusions \label{sect:conclusions}}
The SPHERE-2 detector which operated in 2008--2013 had a large array of supplementary sensors that allowed to control and later reconstruct the state of the detector and measurement conditions. 

For the reflected CL method the information on detector position and orientation is vital. Their values were measured with good precision and reliability and were cross-checked using experimental data.

Measurements of air pressure and temperature during flights gave information on the atmosphere state that will allow to introduce different atmospheric models into analysis and to account their strong impact on the results. Availability of the data on the atmosphere state at the moment of the EAS detection allows higher primary mass reconstruction.

\section{Acknowledgments}

We are grateful to the Lebedev Physical Institute of the Russian Academy of Sciences group (leader S.B.~Shaulov) for assistance in assembling and testing the electronic equipment and in preparation of expeditions. We also thank the Baikal-GVD collaboration and G.V.~Domogatsky (Institute for Nuclear Research, Russian Academy of Sciences) for the support of the SPHERE experiment at the Baikal Lake scientific station.
M. Finger and M. Finger Jr. were supported by MEYS of Czech Republic grants LG14004 and LG18022.

\section{References}

\bibliography{Sphere-Data.bib}

\end{document}